\documentclass[useAMS,usenatbib]{mnras}                                                                           \usepackage[pdftex]{graphicx}
\usepackage{epstopdf}
\usepackage{color}
\usepackage[utf8]{inputenc}
\newif\ifAMStwofonts

\def\degr{\hbox{$^\circ$}}
\def\arcmin{\hbox{$^\prime$}}

\def\utw{\smash{\rlap{\lower5pt\hbox{$\sim$}}}}
\def\udtw{\smash{\rlap{\lower6pt\hbox{$\approx$}}}}

\def\loa{\mathrel{\mathchoice {\vcenter{\offinterlineskip
\def\reference{\parskip 0pt\par\noindent\hangindent 0.5 truecm}
\def\kms{km ${\rm s}^{-1}$}
\halign{\hfil$\displaystyle##$\hfil\cr<\cr\approx\cr}}}
{\vcenter{\offinterlineskip\halign{\hfil$\textstyle##$\hfil\cr
<\cr\approx\cr}}}
{\vcenter{\offinterlineskip\halign{\hfil$\scriptstyle##$\hfil\cr
<\cr\approx\cr}}}
{\vcenter{\offinterlineskip\halign{\hfil$\scriptscriptstyle##$\hfil\cr
<\cr\approx\cr}}}}}

\begin{document}

\title[Absolute Parameters of Young Stars]{Absolute Parameters of Young Stars: V Puppis}

\author[E.\ Budding et al.]{E.\ Budding$^{1,2,3,4}$, T.\ Love$^{1,4}$, M.\ G.\ Blackford$^{4}$,  
 T.\ Banks$^{5,6}$, \and M.\ J.\ Rhodes$^{7}$  
\vspace{2ex}\\
$^{1}$ Visiting astronomer, University of Canterbury Mt John Observatory, University of Canterbury, Private Bag 4800, Christchurch 8140, NZ; \\   
$^{2}$ Carter Observatory, Wellington 6012, NZ;\\
$^{3}$ School of Chemical \& Physical Sciences, Victoria University of Wellington, Wellington 6012, New Zealand.\\
$^{4}$ Variable Stars South, RASNZ; \\ 
$^{5}$ Nielsen, Data Science, 200 W Jackson Blvd 17, Chicago, IL 60606, USA. \\
$^{6}$ Physics \& Astronomy, Harper College, 1200 W Algonquin Rd, Palatine, IL 60067, USA \\
$^{7}$ Brigham Young University`, Provo, Utah, USA.\\
}

 
\maketitle


\begin{abstract}
New spectrometric data on V Pup are combined with satellite photometry
(HIPPARCOS and recent TESS) to allow a revision of the 
absolute parameters with increased precision.
We find: 
 $M_1$ =  14.0$\pm$0.5, $M_2$ = 7.3$\pm$0.3  (M$_\odot$); 
$R_{1}$ = 5.48$\pm$0.18, $R_2$ =  4.59$\pm$0.15 (R$_\odot$);
$T_{1}$ 26000$\pm 1000$,  $T_2$ 24000 $\pm$1000 (K), age 5 $\pm$1 (Myr),
photometric distance 320 $\pm$10 (pc). 
The TESS photometry reveals low-amplitude ($\sim$0.002 mag) variations
of  the $\beta$ Cep kind, consistent with the deduced evolutionary condition and age
of the optical primary.  This fact provides independent support to our understanding of 
the system as in a  process of Case A type interactive evolution
that can be compared with  $\mu^1$ Sco.  
  The $\sim$10 M$_{\odot}$ amount of matter shed by the over-luminous present
  secondary must have been mostly ejected from the system rather than 
  transferred, thus taking angular momentum
  out of the orbit and keeping the pair in relative close proximity.   
 New times of minima for V Pup have been studied and the results
 compared with previous analyses. The implied variation of period
 is consistent with the Case A evolutionary model, though we
offer only a tentative sketch of the original arrangement of this massive system.
We are not able to confirm the previously reported  cyclical 
variations having a 5.47 yr period with the new data, though 
a direct comparison between the HIPPARCOS and TESS photometry
points to the presence of third light from a star that is 
cooler than those of the close binary, as mentioned in previous
literature.  

\end{abstract}

\label{firstpage}

\begin{keywords} 
stars: binaries (including multiple) close --- stars: early type ---
stars: variable $\beta$ Cep type ---
stars: individual V Pup
\end{keywords}

\section{Introduction}
Studies of massive young stars, forming part of a `Southern Binaries Programme', were
presented by members of our group in a number of previous papers.  
General background  was given by Budding (2008) and in the
 summary of Idaczyk et al.\ (2013).  For a recent example
see Blackford et al.\ (2019).

Binary stars are still the main source of fundamental data on stellar masses and radii.  
Our access to such advanced facilities as the high-precision photometry
from the TESS satellite (Section~2) and
the high-resolution HERCULES spectrometer (Section~3) for this application enables 
continued progress in refining our knowledge of the properties of stars.  
Popper's (1980) contribution to this subject set accuracy limits of 
a few percent, that was refined to an estimated $\sim$2\% for the 45 
examples studied by Andersen (1991). These review papers report 
an ongoing development of our understanding of stellar astrophysics, 
on the basis of data that has been confirmed to be reliable and of high 
quality.  Andersen et al. (1993) also drew attention to the 
interesting role of close young binaries in understanding the 
relationship of stellar properties to their galactic environment.  
As well, Rucinski (2006) pointed out the observational
neglect of binaries with declinations south of declination  --15$^{\circ}$.  
Such discussion underlies and provides a springboard for the work 
presented here.  

In this paper,
we re-examine the early type eclipsing binary  V Pup.
The system consists of at least two young stars in a very close orbit,
apparently in a `Case A' process of interactive binary evolution.
 V Pup has attracted recent attention  regarding the possibility 
 that it may contain a black hole companion associated with its known
 high energy and microwave emissions (Giacconi et al.,
1974; Groote et al., 1978; Bahcall et al., 1975; Qian et al., 2008; 
Maccarone et al., 2009).
 The issue may be probed more fully  
with the aid of up-to-date measurement and data-analysis procedures 
as well as the observational facilities mentioned above. 

\subsection{V Puppis}

V Pup ( = HD 65818, HIP 38597, WDS  J07582-4915A, HR 3129; Gaia DR2 55171716782683628800)
is a bright ($V = 4.41$; $B - V = -0.17$  $ U - B = -0.96$; types B1Vp + B2)
eclipsing variable of the `EB' type, discovered by   Williams (1886).
It is located on the sky about 2.5$^{\circ}$ south-west of, and 
at a comparable distance ($\sim$300 pc) to, the massive
  multiple star $\gamma^2$ Vel.  Early history of observations
  of the system is briefly reviewed in the paper of Cousins (1947).
  
  There has been ongoing literature discussion about the absolute parameters of V Pup,
  with effects attributed to the presence of gas streams or mass transfer
  distorting radial velocity (RV) data on the stellar components
  (the `Struve-Sahade effect' ---
  see e.g.\  Maury, 1920; Popper, 1947; Frieboes, 1962;  York et al., 1976; 
  Cester et al., 1977; Eaton, 1978; Koch et al., 1981,   Andersen et al., 1983,
   Stickland et al., 1998). 
  The spectral types were given as B1 V and B3 V by Freiboes (1962)
  with corresponding masses  of about 17 and 9  M$_{\odot}$ (Popper, 1980). 
  Andersen et al.\ (1983) revised the parametrization of the system
  with improved data and analysis procedures, dropping the dwarf luminosity
  (V) of the secondary, which could now be assigned a subgiant (IV) classification.
  Their masses were reduced somewhat from those used by Popper to 14.8
   and 7.8   M$_{\odot}$,
  with error estimates of about 2\% of these values.
  The orbital period is relatively short at 1.4545 d.
  Stickland et al.\ (1998) obtained noticeably lower masses from IUE spectrograms,
  though unfortunately, the coverage at one of the elongations was rather incomplete.
  
V Pup appears to have at least one physical companion, associated with the double star
h4025, the other star being an 11th mag object at a separation of about 6 arcsec. 
Two other fainter companions have also been linked to the massive early-type close binary.
York et al.\ (1976), using UV data from the Copernicus satellite, deduced the presence
of an HII region around the binary with an angular extent of $\sim$3.5 arcmin.

This hot young
 binary may have some physical resemblance to $\mu^1$ Sco (Budding et al., 2015)
in showing a semi-detached, near-contact configuration
(cf.\ Schneider et al., 1979; Andersen et al.\ 1983; Bell et al.\ 1987a, 1987b;
Terrell et al.\ 2005), and with this type of hot massive binary this
is often associated with Case A evolution, in which mass loss
from the erstwhile primary continues while the star is still 
on the hydrogen-burning Main Sequence (Sybesma, 1986;
Yakut \& Eggleton, 2005).
   
With its stable light curve and well defined eclipses, 
 timings of the light minima of V Pup 
 may allow the stability of the binary orbit to be checked
 by use of the `O -- C' (observed minus calculated) diagram. 
 Periodic residuals in such data  (Kreiner et al., 2000) have suggested the presence of a 
$\sim$10 M$_{\odot}$ unseen companion to the binary, with
an orbital period of about 5 years (Qian, Liao \& 
Fernandez-Lajus, 2008). Absence of visible indications of such a massive component
in the optical region lead to the idea that this component may
be a black hole.

 Qian et al.\ (2008) argued that the X-ray emission from 
 V Pup is consistent with accretion of stellar winds from the hot binary
 towards such a black hole.  Interestingly, its  mass could then be
  determinable from the observed orbital effects,
  and independently of evolution scenarios for the overall system.  

Triple systems
with accreting components have been invoked in the past, for example, to
provide a physically plausible scenario for the Cygnus X-1 system (Bahcall et al.\ 1975).
Also, the occurrence of long additional periods in systems
with accreting neutron stars and black holes (Priedhorsky \& Terrell, 1983; Gies \& Bolton 1984; Zdziarski et al. 2007) may be explained in this way.
Longer timescale effects in the   
short period neutron-star binary 4U 2129+47 were attributed to an observed F type
wide companion 
(Garcia et al. 1989; Nowak, Heinz \& Begelman 2002), further substantiating
such an hierarchical stellar system model.  Maccarone et al.\ (2009) made
observations aimed at testing the hypothesis that the X-ray 
emission from V Pup is from accretion onto the putative
black hole by searching for radio emission,
since a high ratio of radio to X-ray flux 
should be a characteristic signature of accreting black holes of stellar mass.
They invoked an X-ray image of the field using Chandra, thus confirming
that the X-rays are indeed from V Puppis or its close ($\sim$0.5 arcsec) vicinity.

Maccarone et al.\ gave an upper limit on the
radio flux of about 300 $\mu$Jy.  V Pup was 
identified in its X-ray emission with a luminosity of about 3$\times 10^{31}$
 erg/sec.  This  value is much
lower than what had been reported in low angular resolution surveys of
the past. Maccarone et al.'s results 
were interpreted to show  that the X-ray emission
comes from mass transfer between the two B stars in the system.
The possibility of X-ray emission from the black hole
accreting stellar wind from one or both of the B stars could not be ruled out, however.
 
In fact, massive early type stars, such as those in the V Pup system, 
are known to produce strong winds driven by the radiation pressure from the 
high temperature stellar surfaces.  In a binary system consisting of two such 
stars the winds will interact, producing exotic shock effects observed across 
the spectrum, but noticeably in the X-ray range 
(e.g. Pollock, 1987; Chlebowski \& Garmany, 1991; Corcoran, 1996) and as synchrotron
 radio-emission.   Such emissions tend to have a complex, intermittent 
 or highly variable behaviour, particularly in higher energy ranges, 
 and observations do not always support the predictions of basic models 
 (e.g. Prilutski \& Usov, 1976; Cherepashchuk, 1976; Stevens et al. 1992).  
 As more data have been collected over the years, fuller recognition of these 
 anomalies has grown and theory developed to introduce more refined models
   (Pittard, 2009;  Parkin et al. 2011; Rauw \& Naz\'{e}, 2016). It seems likely 
   that the high energy emissions noted for V Pup, as well as their variability, 
   may be accounted for by such modelling, without requiring the more extreme 
   configuration invoked by Qian et al.\ (2008).

 In the next section, we present and analyse photometry of V Pup. We follow this 
with some new spectroscopic data in Section~3. Combination of the
results from the photometric and spectroscopic analyses lead to
 our derived set of absolute parameters in Section~4, and we discuss O -- C analysis of the system in Section~5. 
 These findings are discussed and interpreted in terms of 
 relevant stellar astrophysical models in the Section~6.

\section{Photometry}
 
\begin{figure}
\label{fig-1}
\centering
\includegraphics[height=6cm]{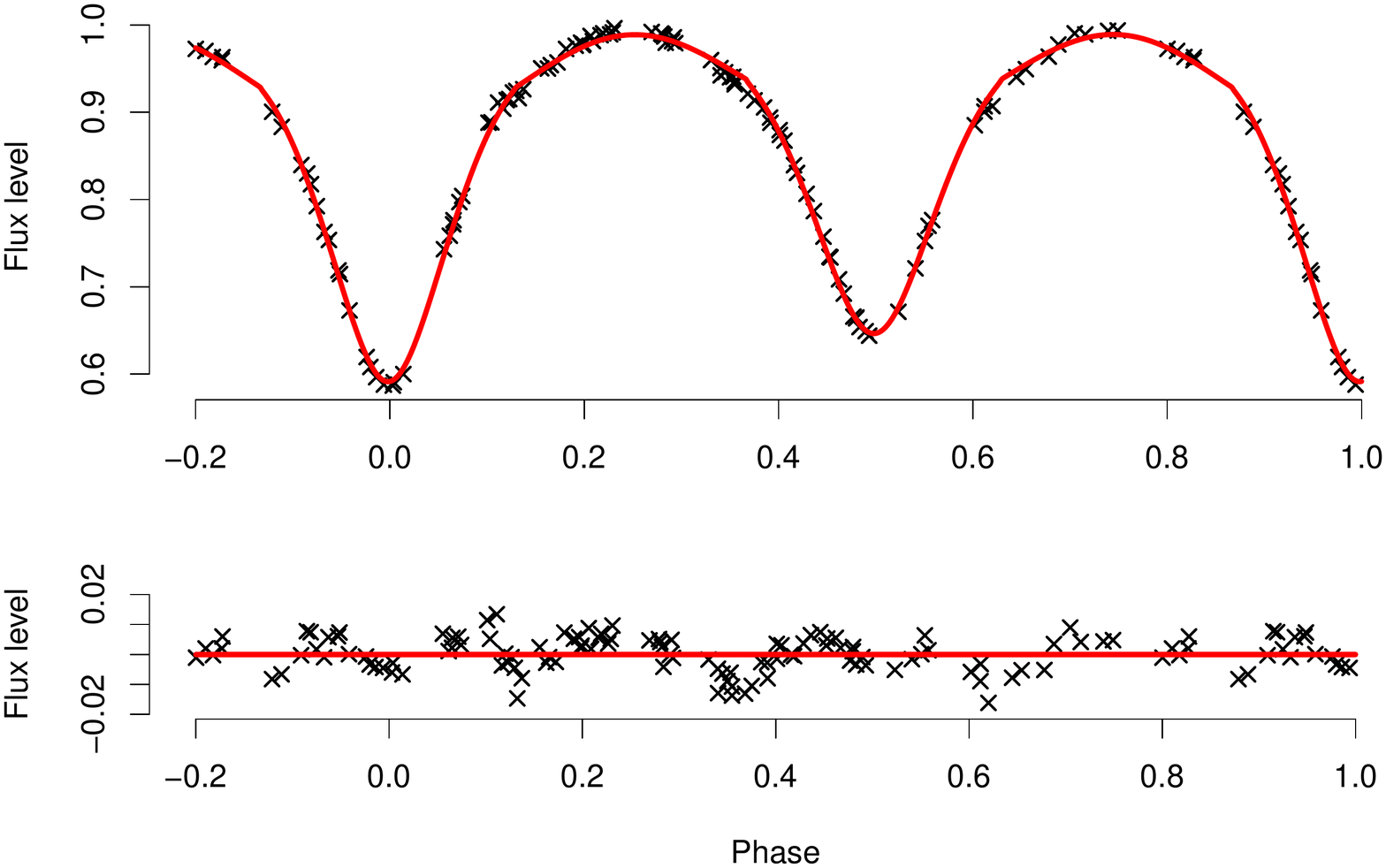}
\caption{HIPPARCOS photometry of V Pup with Radau-model fitting.
Residuals (crosses) are shown distributed about the lower axis.}
\end{figure}

 As before in this programme we 
 looked into previous photometry, particularly that of the 
 HIPPARCOS satellite (ESA, 1997).
The light-curve has been phased in the HIPPARCOS Epoch Photometry Annex according to the
ephemeris Min I = 2448500.595 + 1.454507E. 
Since the data were gathered over a $\sim$3 year interval after JD 2447862.0708 
 a more appropriate epoch for the HIPPARCOS coverage would be 
 HJD 2448455.5053; some 31 light cycles (45.090 d) earlier
 and close to the mean date of the data series.
We adopted this for subsequent reference purposes.

A preliminary fitting to the HIPPARCOS light curve is shown in Fig~1
and a corresponding set of parameters listed in Table~1.
The mass ratio was fixed at 0.55 from initial radial velocity findings, 
discussed in the next section.
Effective photospheric temperatures could be also 
estimated from the spectroscopy with the values shown
in Table~1.  Formal error estimates are provided by the fitting program
{\sc WinFitter} that follow from the principles set out in 
 Bevington (1969), chapter 11.  The procedure involves numerically 
inverting the determinacy Hessian of the $\chi^2$ variate in the vicinity of
its minimum.  This Hessian matrix should be positive definite for a properly posed
inversion problem.  The resulting errors then include the effects of inter-correlations
between the parameters selected for adjustment (see also Banks \& Budding, 1990,
for further background on this topic). 

We can at once deduce, from the comparability of the luminosities
despite the appreciable difference in masses, that we are not dealing with a Main Sequence pair.   
Kopal's (1959, Table~3.3) approximation that $r_1 + r_2 \approx 0.75$ for 
contact binaries, indicates that these stars
must be relatively close to contact at the internal Lagrangian point L$_1$.
Indeed, the relative size of the secondary points to a filling of its `Roche lobe'.
This point, together with the closeness of the stars, confirms that
this eclipsing binary is of the semi-detached Case-A-evolution type that may be compared with 
$\mu^1$ Sco (cf.\ Introduction). 

A small shift from zero in the phase of minimum light $\Delta \phi_0$ is given in 
Table~1. The movement is in the sense that the observed minimum was late from
 its predicted time by an amount of some 0.0019 in phase, or 0.0028 days from the predicted time
using the HIPPARCOS ephemeris given above.  We deduce that the time of observed minimum 
should be 24448455.5081, or some 13618.01536 cycles after Kreiner's epoch
that is used as a basis for O --C studies.
This leads to an O -- C value of +0.0223 d, which is slightly greater than
the +0.0198 d given by Qian et al.\ (2008).

\begin{table}
\begin{center}
\caption{Curve fitting results for archival HIPPARCOS 
photometry of V Pup using {\sc WinFitter 6.2}.
Parameters for which no error estimate is given are adopted from
information separate to the fitting.
The HIPPARCOS raw data were phased according to the ephemeris
provided in the HIPPARCOS Photometry Annex (ESA, 1997; -- see text).}
\begin{tabular}{lcc}
   \hline  \\
\multicolumn{1}{c}{Parameter}  & \multicolumn{1}{c}{Value} & 
\multicolumn{1}{c}{Error}\\
\hline \\
$M_2/M_1$ & 0.55 & \\
$L_1$ & 0.58 & 0.01 \\
$L_2$ & 0.42 & 0.01 \\
$r_1$ (mean) & 0.406 & 0.002\\
$r_2$ (mean) & 0.349 & 0.005 \\
$i$ (deg) & 75.1 & 0.4 \\
$T_h$ (K) & 26000 & \\
$T_c$ (K) & 24000 & \\
$u_1$ & 0.28 &\\
$u_2$ & 0.30 &\\
$\Delta\phi_0$(deg) &0.7 & 0.2 \\
$\chi^2/\nu$ & 1.02 & \\
$\Delta l $& 0.006  &\\
\hline
\end{tabular}
\end{center}
\end{table}
 
 \subsection{TESS photometry}
The Transiting Exoplanet Survey Satellite (TESS) was launched in April 2018, and began regular science
operations by the end of July in that year.  This NASA-supported mission is primarily aimed at
a near complete survey of the whole sky, searching for exoplanet transiting systems
(Ricker et al., 2015).
In its first two years  TESS should closely monitor over
   200,000 Main Sequence stars with four wide-field optical CCD cameras.
    Photometry of the target stars is recorded with a 2 minute cadence. 
Additional images are taken of a wide ($\sim 2200$ sq.\ deg.) field every 30 minutes.  
    It is planned to shorten the cadences to 20 sec and 10 minutes in a later extension
    to the mission.

 The data selected for the present study concerns about 15130 individual PDCSAP flux
 measures with the TESS identification of V Pup as 
 tess2019006130736-s0007-0000000269562415-0131-s{\_}lc.  These data, from sector~7, run 
 continuously in 2-minute samples from days BJD
 2458492.628 to 2458516.088, apart from a gap between 2458503.039 to 2458504.702. 
Additional  TESS data on V Pup are available from sectors  and 9, 34 and 35.
The normalized PDCSAP fluxes, having a time-range of over 28 complete cycles of the binary
system, were phased with the HIPPARCOS period and an adopted local epoch of BJD 2458492.98218.
 {\sc WinFitter} was then used to experiment with photometric modelling of these data.
 
 Although the TESS data are of very high accuracy, enabling a remarkably close 
 modelling fit (Fig~2), 
 at an initial stage it became clear that there were small short-term complications in the 
 form of the data compared with expectation for a standard binary system pattern of light variation. 
The complete data-set was then binned by a factor of 10 and an optimal model sought, using the
results of Table~1 as a guide.  The adopted model parameters are listed in Table~2.
The fitted light curve is shown in Fig~2, and
Fig~3 displays a corresponding illustration of the basic stellar model.

\begin{figure}
\label{fig-2}
\centering
\includegraphics[height=6cm]{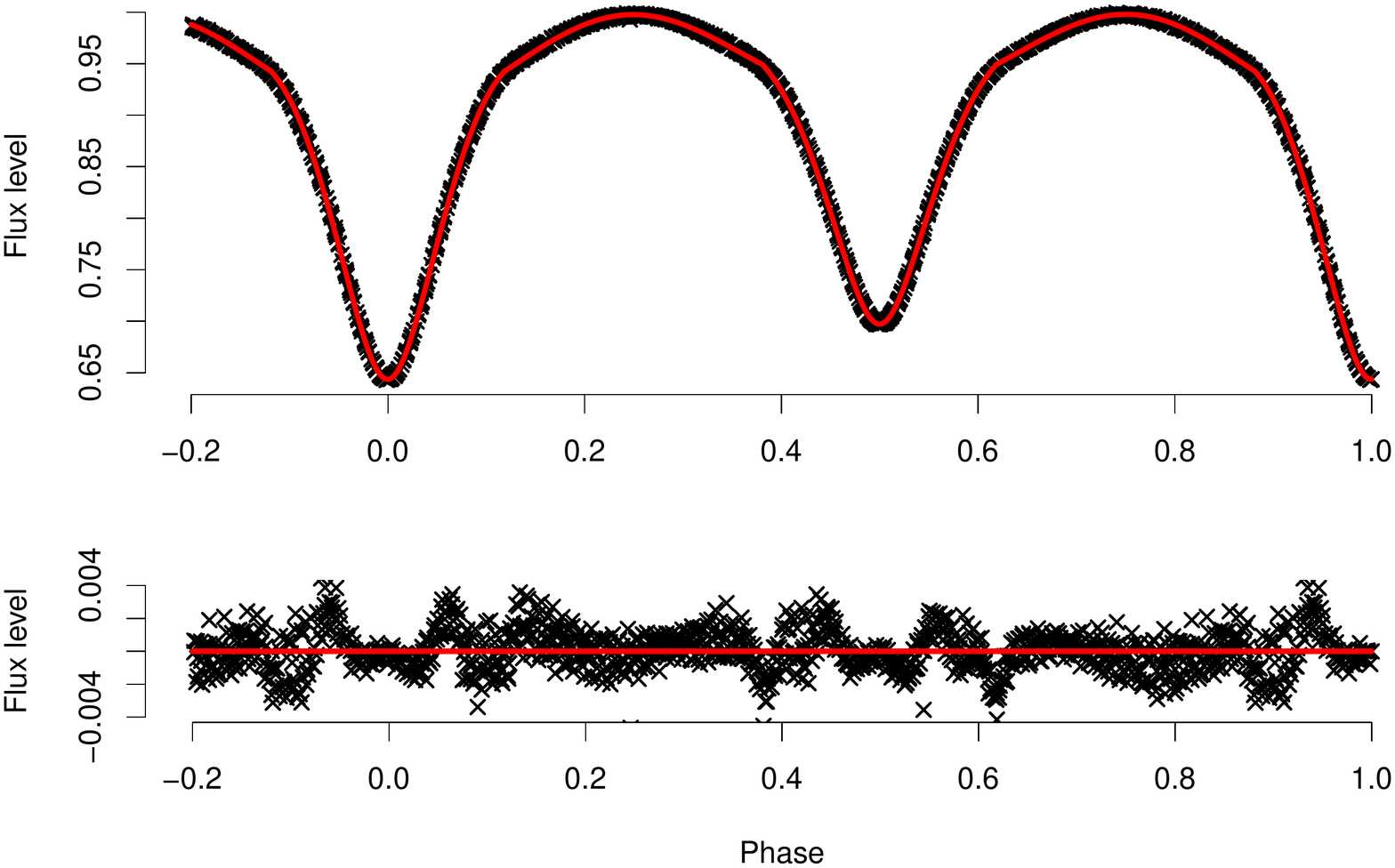}
\caption{{\sc WinFitter} model fit to the complete TESS photometric data-set for V Pup, binned by a factor of 10 down to 1513 individual values.  The scatter of individual readings is of order 0.0002,
but systematic departures of up to an order of magnitude greater than that can be seen in the 
residuals in the regions of the eclipses when
plotted about the mean flux level 0.55 in this diagram.  This is mostly accounted for 
by the fitting function's quadratic approximation being  inadequate to represent the 
stars' photospheric limb-darkening effect with sufficient accuracy
of detail.  The parameters corresponding to this fitting are given in Table~2.
 }
\end{figure} 

These parameters have some noticeable differences from the ones in Table~1.
But the forms of the light curves are also seen to be different: while both light curves
show the peak brightness at close to unity and the minima are in about the
same proportions, those in Fig~2 are more than 10\% 
shallower than those of Fig~1.  This is probably mainly due to the difference in
effective wavelengths of the HIPPARCOS and TESS photometers.
The TESS detector bandpass spans from 600 --- 1000 nm and is centered on the
Cousins I-band (7865 \AA), but with the high temperature expected for the source
($\sim$25000 K -- Table~2), the effective wavelength works out at close to 731 nm. 
On the other hand, 
 the HIPPARCOS filter centres at 5593 \AA\ 
 with a corresponding effective wavelength of $\sim$549 nm (Mann \& von Braun, 2015).

 The  possible contribution of an additional cooler component, or components,
 therefore arises (cf.\ Eaton, 1978).
While there is no question that the TESS data have a higher measurement 
precision than those of HIPPARCOS, (by a couple of orders of magnitude, to judge by
the scatter of individual measures), the TESS pixels (21 arcsec in extent) 
include significantly more background sky than the field of HIPPARCOS. 
The WDS (Worley \& Douglass, 1997) lists four visual companions to
V Pup, but none of the nearest 3 within 20 arcsec are brighter than mag 11.5,
while HJ 4025D, the brightest at V = 9.88 and separated at 40 arcsec,
can not have added significantly to background illumination,
though the possibility of error in the background subtraction procedure,
given the large pixels of TESS, cannot be ruled out.
Table~2 presents this as $L_3$: the
additional light at the longer wavelength that reduces the apparent amplitude of 
the variation due to the close binary alone.

The residuals from the fitting to the binned complete data-set
shown in Fig~2, whilst evidently less than those of Fig~1,
reveal small systematic features, of amplitude $\loa 0.002$, that affect the eclipse phase ranges.
These are associated mainly with shortcomings of the quadratic formula employed 
for the limb-darkening of the stars.  Notwithstanding the eclipse regions, residuals in 
the out-of-eclipse phases  are reasonably flat and close to zero.

When we examine individual light curves, however, 
we observe an additional intrinsic variation (see Figs~ 4 and 5)
with a quasi-period of around 0.075 that of the system, i.e.\ $\approx 2.6$ h, that points to 
the massive B-type primary being a $\beta$ Cep type variable, with  the rather low amplitude of
$\sim$2 mmag in the near IR region of the TESS filter. Anticipating the calibration of the primary's
properties carried out in Section~4, we place that star in the 
H-R diagram in Fig~6, where indeed an incipient $\beta$ Cep variability
is likely in view of the primary star's proximity to the instability strip.
This discovery potentially offers an additional check on the 
system's parametrization. 							

 \begin{figure}
\label{fig-3}
\hspace{-0.5cm}
\includegraphics[height=4cm]{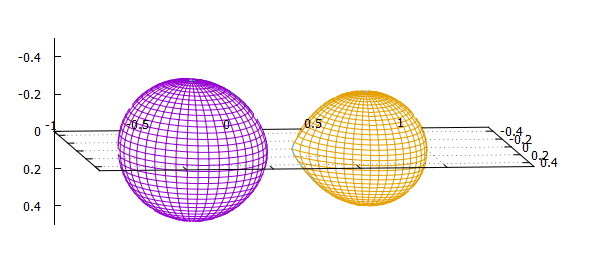}
\caption{{\sc Gnuplot} illustration of V Pup using a standard Roche-model   
corresponding to the
parameters of Table~2.  The secondary is close to contact with its `Roche lobe'. }
\end{figure}

\begin{table}
\begin{center}
\caption{Curve fitting results for TESS
photometry of V Pup using {\sc WinFitter 6.5}.
Parameters for which no error estimate is given are adopted from
information separate to the fitting.}
\begin{tabular}{lcc}
   \hline  \\
\multicolumn{1}{c}{Parameter}  & \multicolumn{1}{c}{Value} & 
\multicolumn{1}{c}{Error}\\
\hline \\
$M_2/M_1$ & 0.50 & \\
$L_1$ & 0.518 & 0.002 \\
$L_2$ & 0.325 & 0.003 \\
$L_3$ & 0.162 &  0.004 \\
$r_1$ (mean) & 0.366 & 0.002\\
$r_2$ (mean) & 0.307 & 0.002 \\
$i$ (deg) & 80.5 & 0.3 \\
$T_h$ (K) & 26000 & \\
$T_c$ (K) & 24000 & \\
$u_{11}$ & 0.25 & 0.01\\
$u_{12}$ & 0.19 & 0.01\\
$u_{21}$ & --0.03 &\\
$u_{22}$ & --0.03 &\\
$\Delta\phi_0$(deg) &--2.04 & 0.02 \\
$\chi^2/\nu$ & 0.86 & \\
$\Delta l $& 0.0015  &\\
\hline
\end{tabular}
\end{center}
\end{table}

\begin{figure}
\label{fig-4}
\includegraphics[height=9cm]{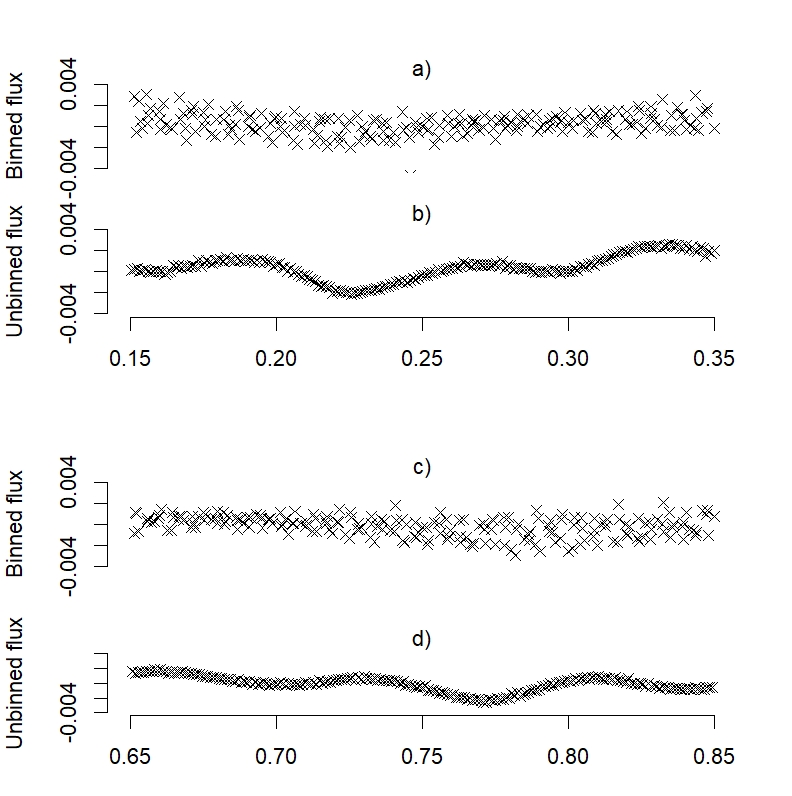}
\caption{Panels (a) and (c) show the light-level of the out-of-eclipse phases
resulting from the binning of the 14+ individual TESS light curves of V Pup.
This appears to have mostly smoothed out the intrinsic $\beta$ Cep variation 
visible in individual light curves, as shown in panels (b) and (d)
(although traces of some predominating low-amplitude tendency can be seen in the binned data).
}
\end{figure}
 
 \begin{figure}
\label{fig-5}
\includegraphics[height=6cm]{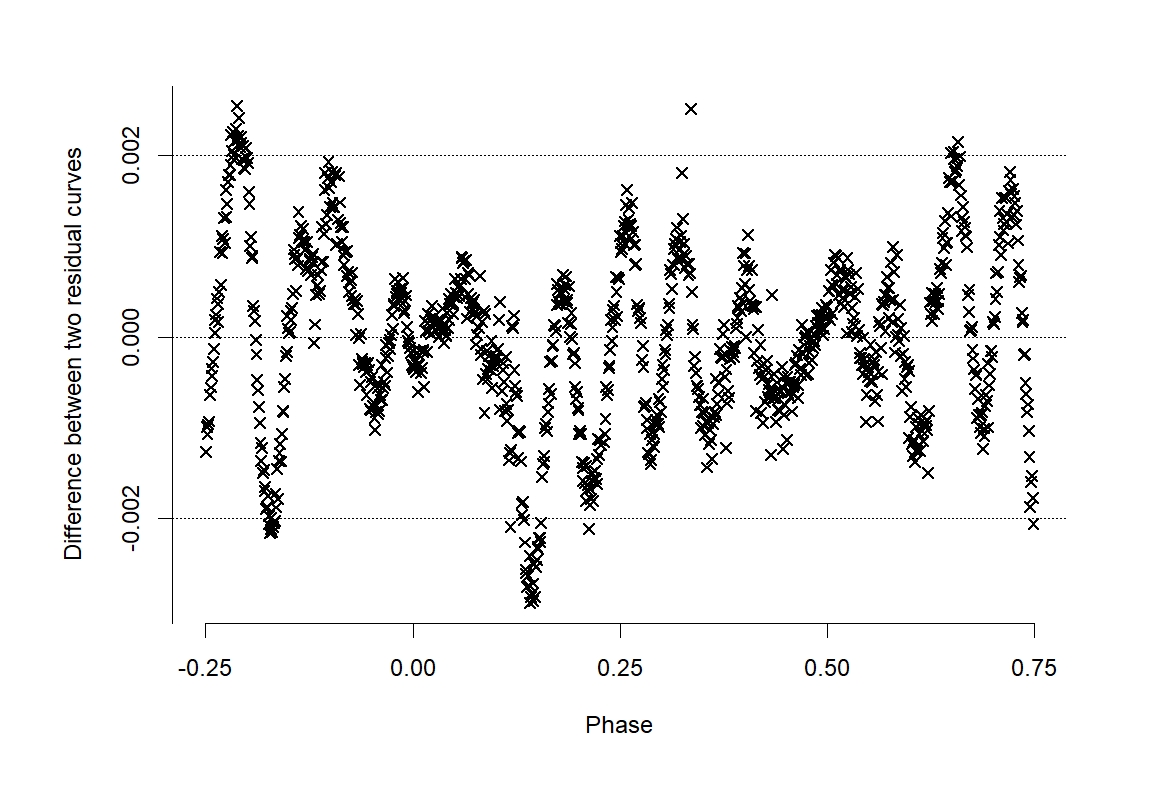}
\caption{The character of the short-term oscillatory behaviour is confirmed by
plotting the difference between the residuals from the single light curve fit and 
those of the 14 combined and binned light curves.  The apparent damping out
of the oscillator behaviour through the primary eclipse is indicative that it is
the primary star which undergoes the pulsations, but this evidence, by itself,
is not conclusive. }
\end{figure}
 
\begin{figure}
\label{fig-6}
\hspace{-3cm}
\includegraphics[height=10cm]{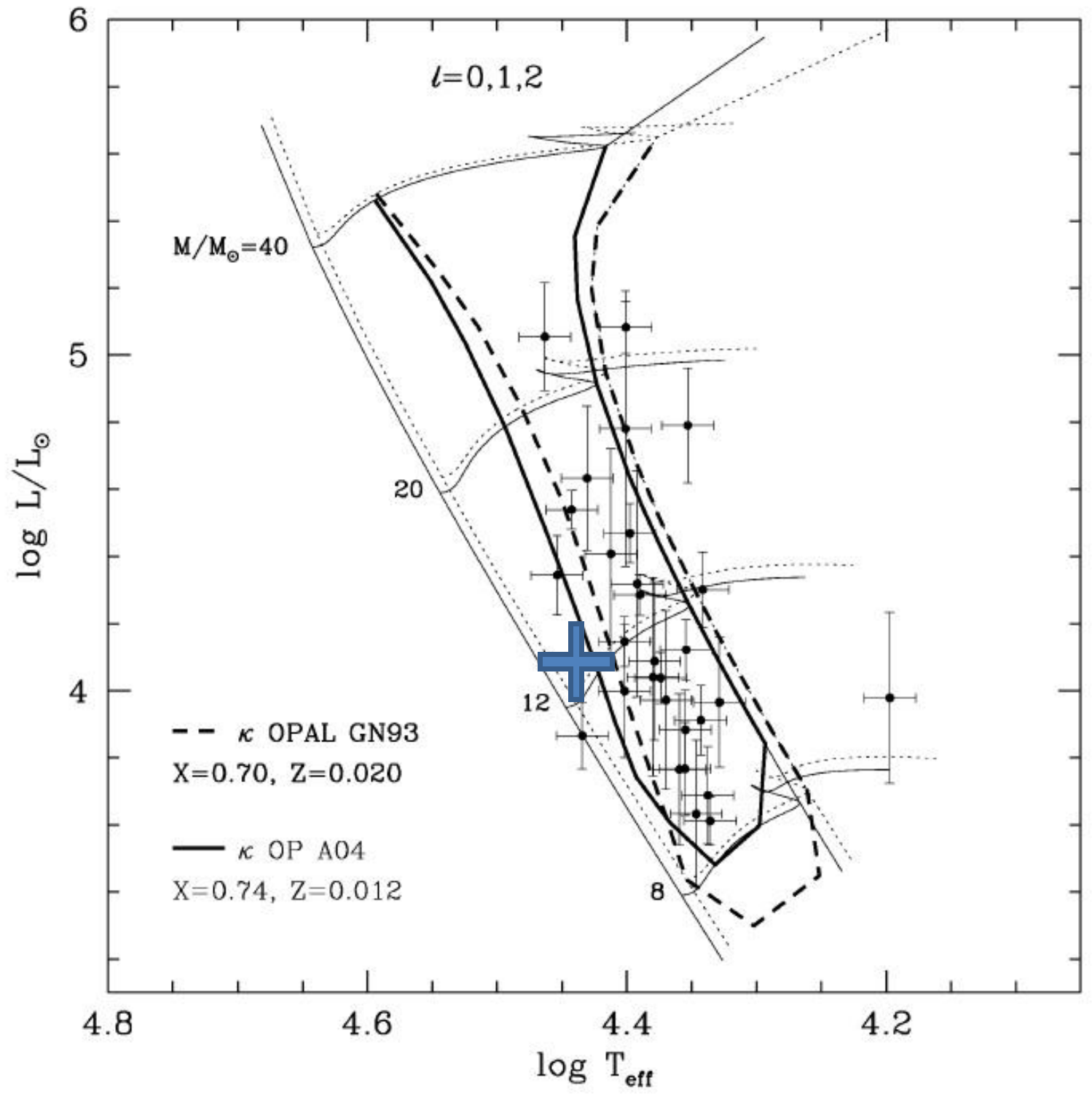}
\vspace{-1cm}
\caption{$\beta$ Cep stars in the HR diagram.
The revised instability strip (copied from Fig~2 in Pamyatnykh, 2007),
that uses improved OP opacities, is delineated by the 
thick-line border.  The dense blue cross locates the primary of V Pup as an incipient
$\beta$ Cep type variable in the H-R diagram.}
\end{figure}

\begin{figure}
\label{fig-7}
\hspace{-0.5cm}
\includegraphics[height=6.5cm]{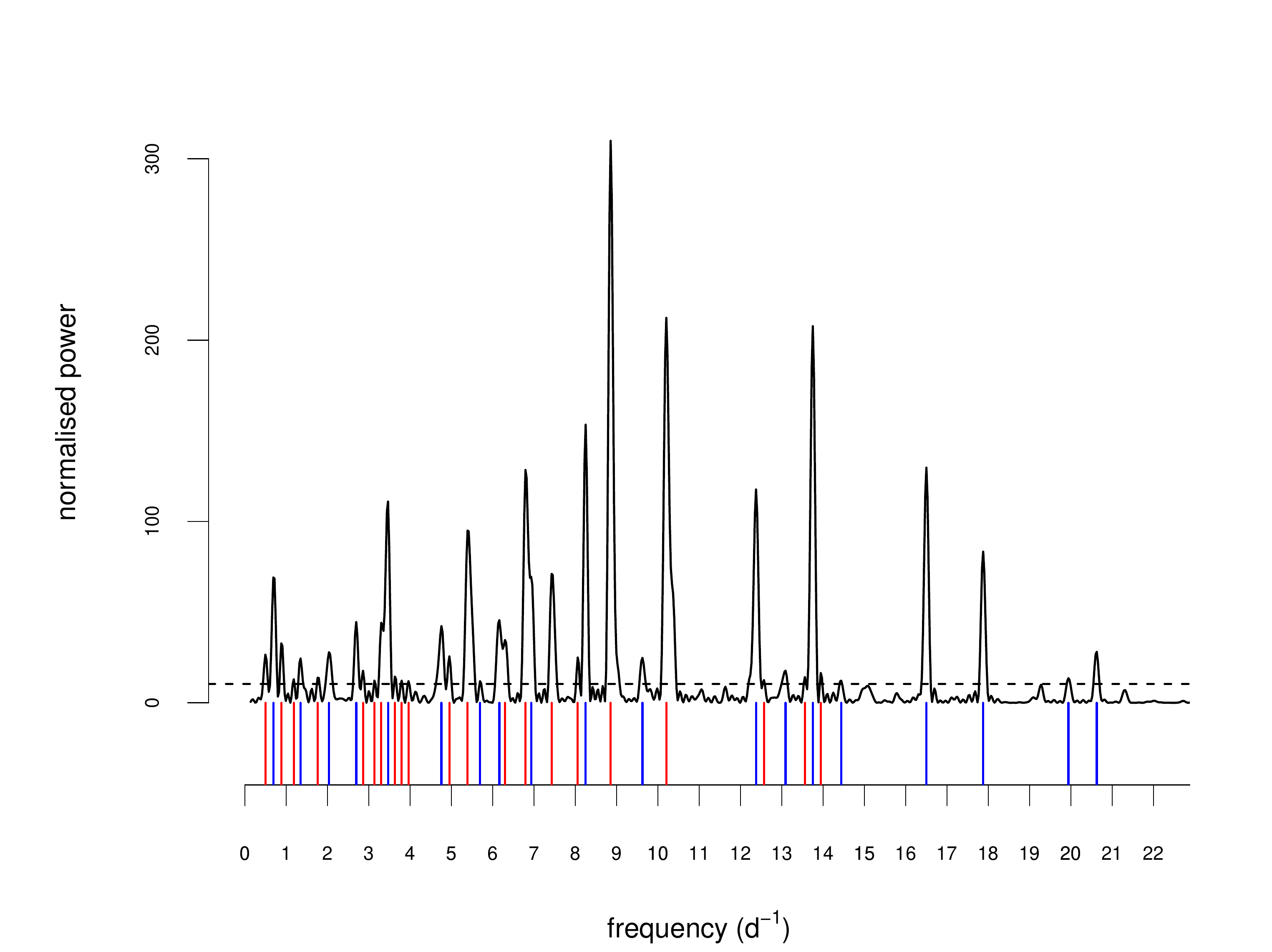}
\caption{ Lomb-Scargle periodogram of pulsational effects in the V Pup system.
This may include a possible $\beta$-Cep type behaviour of the primary star. 
Solid blue lines indicate  significant (p=0.01) frequencies that are within 0.1 
of an integer multiple of the orbital frequency. Other significant 
frequencies are indicated by red lines.  
The two strongest signals are at 8.8555 d$^{-1}$ (= 2.71 h) and 10.2030 d$^{-1}$ (=2.35 h).
}
\end{figure}

  Lomb-Scargle analysis (cf.\ Ruf, 1999) 
of the flux residuals reveals a complex pattern
(Fig~7), with some 40 identified frequencies significant at the 99\% confidence level.
Nineteen of these are close to multiples of the orbital frequency, 
and therefore suspected to be aliases or artefacts from the close binary 
light curve modelling.  In this connection, complications also
arise from inadequacy of the limb-darkening series approximation.  This is known to
introduce additional oscillatory effects in the residuals from the
eclipse phases.  
The {\sc Period04} program (Lenz \& Breger 2005) was also used to check on 
the pulsations. Its pre-whitening technique was applied 
to data outside the two eclipses, from which some 13 significant
frequencies were found, supporting the main clump of strong
contributions in the $\sim$ $P$/5 -- $P$/10 range.
Further work is
clearly needed to investigate and quantify the pulsation spectrum and its source. 
 
\section{Spectroscopic data}
 
Spectroscopic data were mostly gathered using the
High Efficiency and Resolution Canterbury University Large
\'{E}chelle Spectrograph (HERCULES) of the Department of
Physics and Astronomy, University of Canterbury, New
Zealand (Hearnshaw et al., 2002).
 This was attached to the 1m McLellan telescope at
the University of Canterbury Mt John Observatory (UCMJO) near Lake
Tekapo ($\sim$ 43{\degr}59{\arcmin}S, 174{\degr}27{\arcmin}E).  Over 120 spectra 
of V Pup were collected during
fairly clear weather during the period Feb 5-12, 2020.

The phase coverage of these UCMJO observations was about 65\% of the complete cycle.
Uncovered phases have been observed subsequently (TL) from the Saesteorra Observatory in the Wairarapa region of New Zealand 
(41$^{\circ}$ 13$^{\prime}$ 07$^{{\prime}{\prime}}$S;  
175$^{\circ}$ 27$^{\prime}$ 51$^{{\prime}{\prime}}$
E) with a 12 inch f/8 Ritchey-Chretien telescope.  The spectroscopic equipment there
consisted of a Lhires Littrow design spectrograph (Thizy, 2007), 
having a 35 micron slit and a 2400 lines/mm grating.  This has a nominal resolution of 
approximately R=13400. 
 The spectral coverage was set to be from 6607 to 6813 {\AA}, 
so as to encompass the He6678 line. 

The UCMJO spectral images were collected with a
 4k$\times$4k Spectral Instruments (SITe) camera (Skuljan, 2004).
The 100 $\mu$ fibre, which is suited to typical
seeing conditions at Mt John, enables a theoretical resolution of $\sim$40000.
The regular image sequencing for wavelength and relative flux calibration was followed
(cf.\ Blackford et al., 2019), however,
an error in the detector's flux meter led to under-exposure 
of some calibration frames.  Subsequent checks on telluric lines in the science 
frames provided assurance that our final data reductions were accurate and consistent.
Exposure times were usually $\sim$100 seconds for this 4th mag star. 
 The raw observations were reduced using an updated version of the software package {\sc hrsp}
(Skuljan, 2012), that produces wavelength calibrated and normalized
 output conveniently in {\sc fits} formatted files.

 Some 45 clear orders of the \'{e}chelle were set up for inspection. 
These have been studied using  {\sc iraf} software. 
 One of the {\sc iraf} subroutines ({\sc splot}), for example,
allows image statistics to be checked.  We could determine in this way
a signal to noise ratio (S/N) for continuum pixel regions (away from
flaws or telluric effects) to be usually of order 70 (with a 
$\sim$10\% deterioration from orders 85 to 121: the orders 
examined in this study). 
The spectra were also examined using the software package 
Visual Spec ({\sc vspec}; Desnoux \& Buil, 2005), which
is an MS-Windows$^{\rm TM}$ based package that allows convenient analysis of spectra, 
including line fitting and identification functions.

The better-defined He I lines in Table~3 have a depth of
$\sim$5\% of the continuum and a width at base of up to $\sim$10 {\AA} or
$\sim$350 pixels. The lines hardly make a complete separation
 even at
the highest observed elongation. The positioning of a well-formed symmetrical 
line (notably the primary's $\lambda$6678 line)
is consequently expected to be typically achieved to within $\sim$0.15 {\AA}
or (equivalently) $\sim$7 km s$^{-1}$. (The resulting distribution of $\sim$130 measured 
RV displacements for either star is seen in Fig~8, while
the He I $\lambda$6678 feature itself is shown in Fig~9.)

\begin{table}
\begin{center}
\caption{Identified spectral lines for V Pup (p $\equiv$ primary; s $\equiv$ secondary).
\label{tbl-3}}
\begin{tabular}{ccll}
  & & &  \\
\multicolumn{1}{c}{Species}  & \multicolumn{1}{c}{Order no.} &
\multicolumn{1}{l}{Adopted $\lambda$} & \multicolumn{1}{l}{Comment}  \\
He I         & 85        &  6678.149  &   well-defined p \& s \\
H$_{\alpha}$ & 87        &  6562.817  &   strong, complex      \\
He I         & 97       &  5875.650  &  well-defined p \& s \\
C II      (*)   & 110      &  5145.16  & p weak   \\
He I         & 112      &  5047.736  & weak, ill-defined p \& s  \\
He I         & 113      &  5015.675  &  weak\\
He I         & 115      &  4921.929   & strong p \& s   \\
H$_{\beta}$  & 117      &  4861.332  &  strong, complex      \\
He I         & 121      &  4713.146  & weak p \& s \\
He II        & 122      &  4685.7    &  p only \\
C III+ OII   & 122      &  4649.1 &   strong blend   \\
Si III        & 125      &  4574.78 &  edge of order     \\
Si III        & 125      &  4567.87  &  blend \\
Si III        & 125      &  4552.65  & blend     \\
\hline \\
\end{tabular}
\end{center}
\end{table}

{\sc vspec} also allows the equivalent width (W$_\lambda$) of features
to be directly read from  its on-screen spectrometry bar.
We could check, in this way, the spectral type assignations
in the literature.  Direct measures of equivalent widths
have to be scaled against the relative luminosities of the two stars.
The results of measuring a dozen pairs of the well-recorded 
$\lambda$6678 He I feature at elongation are presented in Table~4.

\begin{table}
\begin{center}
\caption{Equivalent width measures of the $\lambda$6678 He I feature
for V Pup at the two elongations.
\label{tbl-4}}
\begin{tabular}{ccccc}
  & & & & \\
\multicolumn{1}{c}{Phase}  & \multicolumn{1}{c}{ Primary} &
\multicolumn{1}{l}{Error} & \multicolumn{1}{l}{Secondary} & \multicolumn{1}{l}{Error} \\
\multicolumn{1}{c}{}  & \multicolumn{1}{c}{ ew (\AA)} &
\multicolumn{1}{c}{\AA} & \multicolumn{1}{c}{ew (\AA)} & \multicolumn{1}{c}{\AA} \\
0.25  &0.64&	0.04&	0.52&	0.08 \\
0.75  &0.58&	0.03&	0.60& 	0.10 \\
\hline \\
\end{tabular}
\end{center}
\end{table}
These results are consistent with the B1V and B3IV literature types (cf.\ Leone \& Lanzafame, 1998;
who provide effective
  temperatures that go with the type assignations),
though the discrimination is not strong  from these He I measures.
  The scatter of W$_\lambda$ values indicates errors  of $\sim$500 K
  in the assigned temperatures.
There is some support for the approaching hemispheres to  show slightly 
stronger absorptions (by $\sim$10\%) as discussed in previous
papers (see Section~1.1).
 
\subsection{Radial velocities}

The RVs listed in Table~5  derived from profile fitting to  
determine the selected He I lines' centre-of-light wavelengths and
the corresponding Doppler displacements.
The well-defined He I line at 6678 has been used mostly,
with additional checks made with the He I 5876 line. 
The He I lines have the best S/N ratios in our spectra, apart from 
the hydrogen lines that are difficult to model due to their highly blended wings.  
Each useful line was fitted separately at least three times and a final RV value obtained by averaging, noting the standard deviations in the same process.   

The RVs in Table~5 have been corrected
to solar system heliocentric values using 
{\sc hrsp} and {\sc vspec} data-reduction tools.  The listed  orbital phases
are derived from  the  ephemeris given by Andersen et al.\ (1983), i.e.\
\begin{equation}
{\rm Min I} \, : \,\, {\rm HJD}2445367.60633 + 1.4544859 \,\, E
\end{equation}
This ephemeris is cited in the GCVS of Khopolov et al.\ (1987),
  and appears to be an accurate representation of early 
  observations of the system.  The adopted period is close to the mean value given by 
  Kreiner et al.\ 2000, though it appears clear from more recent data
  that there is some variation in the period. This will be discussed in Section~5.
  
   Measured RVs
are presented with an accuracy 0.1 km s$^{-1}$, 
but that overestimates the
accuracy of individual points (see above). 
Line-positioning  depends mainly on the
line properties, rather than the available 
spectrographic resolution, which is relatively high for HERCULES.
But while  RV measures are given with one significant
decimal place in Table~5, equivalent to $\sim$0.1 pixel at $\lambda$6678,
it can be seen in Fig~8 that the residuals are, in general,
quite a bit greater than that.   Some of these departures from the model
appear to show systematic trending,
but in the vicinity of the conjunctions blending introduces 
noticeable uncertainty into the measurement.  For this reason
the regions of the RV-curve fitted in order to define the velocity amplitudes in Table~6 
were confined to  phase regions outside the eclipse regions.

\begin{table}
\begin{center}
\caption{Radial velocity data for V Pup.
\label{tbl-5}} 
\begin{footnotesize}
   \begin{tabular}{ccrr} 
\multicolumn{1}{c}{HJD}  & \multicolumn{1}{c}{Orbital} & 
\multicolumn{1}{c}{RV1} & \multicolumn{1}{c}{RV2} \\ 
\multicolumn{1}{c}{2458880+}  & \multicolumn{1}{c}{phase } &
\multicolumn{1}{c}{km s$^{-1}$} & \multicolumn{1}{c}{km s$^{-1}$}\\ 
\hline \\
5.0856&	0.6475&	120.4&	--170.4\\	   
5.0909&	0.6512&	122.5&	--189.5\\	   
5.1434&	0.6873&	154.4&	--225.2\\	   
5.1473&	0.6899&	157.2&	--229.2\\	   
5.8750&	0.1902&	--109.6&	260.3\\	   
5.8804&	0.1940&	--121.1&	260.9\\	   
5.8852&	0.1973&	--109.2&	259.5\\ 	   
5.8897&	0.2004&	--105.1&	267.4\\ 	   
5.8938&	0.2032&	--111.6&	278.0\\ 	   
5.8984&	0.2063&	--109.6&	278.4\\ 	   
5.9034&	0.2098&	--118.4&	284.5\\ 	   
5.9079&	0.2129&	--113.8&	284.3\\ 	   
5.9134&	0.2167&	--119.5&	295.8\\ 	   
5.9176&	0.2195&	--119.9&	306.8\\ 	   
6.0978&	0.3434&	--140.6&	376.0\\ 	   
6.1013&	0.3458&	--134.8&	372.7\\ 	   
6.1050&	0.3484&	--133.8&	361.0\\ 	   
6.1095&	0.3515&	--143.6&	370.4\\ 	   
6.1149&	0.3552&	--132.3&	370.8\\ 	   
6.1195&	0.3584&	--133.0&	362.0\\ 	   
6.1240&	0.3614&	--139.1&	369.6\\ 	   	   
6.8565&	0.8651&	212.5&	--290.5\\ 	   
6.8601&	0.8675&	213.1&	--283.9\\ 	   
6.8655&	0.8712&	210.2&	--280.8\\ 	   
6.8700&	0.8743&	206.4&	--279.0\\ 	   
6.8757&	0.8783&	204.8&	--289.2\\ 	   
6.8808&	0.8818&	200.4&	--292.1\\ 	   
6.9411&	0.9232&	186.6&	--248.8\\ 	   
6.9456&	0.9263&	177.1&	--238.4\\ 	   
6.9495&	0.9290&	184.4&	--232.3\\ 	   
6.9546&	0.9325&	182.6&	--240.7\\ 	   
6.9602&	0.9364&	183.7&	--235.6\\ 	   
7.0098&	0.9705&	161.7&	--163.2\\ 	   
7.0150&	0.9740&	164.8&	--144.5\\ 	   
7.0184&	0.9764&	170.6&	--178.6\\ 	   
7.0219&	0.9788&	158.4&	--154.4\\ 	   
7.0481&	0.9968&	160.1&	--106.6\\ 	   
7.0518&	0.9993&	166.9&	--92.1\\  	   
7.0553&	0.0017&	162.1&	--76.4\\ 	   
7.0591&	0.0044&	163.1&	--76.1\\ 	   
7.0626&	0.0068&	136.3&	--111.8\\ 	   
7.0662&	0.0092&	158.4&	--80.2\\ 	   
7.0716&	0.0129&	150.6&	--80.5\\ 	   
7.0756&	0.0157&	166.6&	--53.2\\ 	   
7.0792&	0.0182&	159.3&	--54.2\\ 	   
7.1155&	0.0431&	135.3&	--11.0\\ 	   
7.8768&	0.5665&	---&	102.3\\ 	   
7.8805&	0.5691&	---&	97.8\\ 	   
7.8859&	0.5728&	---&	103.6\\ 	   
7.8912&	0.5764&	---&	99.1\\ 	   
7.8968&	0.5803&	---&	83.8\\ 	   
7.9603&	0.6240&	---&	--91.8\\ 	   
7.9924&	0.6460&	143.4&	--176.2\\ 	   
7.9998&	0.6511&	149.7&	--- \\ 	
8.0095&	0.6578&	144.8&	--- \\ 	   
8.0002&	0.6650&	129.1&	--203.7\\ 	   
8.0304&	0.6721&	138.4&	--198.4\\ 	   
8.0614&	0.6935&	159.5&	--210.3\\ 	   
8.0721&	0.7008&	165.9&	--239.4\\ 	   
8.0831&	0.7084&	173.9&	--246.9\\ 	   
8.0928&	0.7150&	179.8&	--253.1\\ 	   
\hline \\
\end{tabular}
\end{footnotesize}
\end{center}
\end{table}

\setcounter{table}{4}
\begin{table}
\begin{center}
\caption{Radial velocity data for V Pup (continued).
\label{tbl-5a}}   
\begin{footnotesize}
   \begin{tabular}{ccrr} 
\multicolumn{1}{c}{HJD}  & \multicolumn{1}{c}{Orbital} & 
\multicolumn{1}{c}{RV1} & \multicolumn{1}{c}{RV2}  \\ 
\multicolumn{1}{c}{2458880+}  & \multicolumn{1}{c}{phase } &
\multicolumn{1}{c}{km s$^{-1}$} & \multicolumn{1}{c}{km s$^{-1}$} \\
\hline \\   
8.1036&	0.7225&	183.7&	--260.2\\ 	   
8.1125&	0.7286&	181.4&	--272.4\\	   
8.1218&	0.7350&	179.2&	--265.8\\ 	   
8.8826&	0.2581&	--138.0&	334.3\\ 	   
8.8879&	0.2617&	--137.8&	342.6\\ 	   
8.8919&	0.2645&	--132.9&	355.0\\ 	   
8.8965&	0.2676&	--129.2&	352.5\\ 	   
8.9039&	0.2727&	--139.3&	362.1\\ 	   
8.9082&	0.2757&	--133.6&	360.9\\ 	   
8.9130&	0.2790&	--138.8&	365.1\\ 	   
8.9177&	0.2822&	--138.4&	352.4\\ 	   
8.9226&	0.2856&	--138.7&	358.2\\ 	   
8.9298&	0.2905&	--144.8&	362.8\\ 	   
8.9338&	0.2933&	--144.0&	365.9\\ 	   
8.9450&	0.3010&	--141.3&	368.4\\ 	   
8.9491&	0.3038&	--139.8&	371.2\\ 	   
8.9533&	0.3067&	--145.3&	373.2\\ 	   
8.9603&	0.3115&	--138.6&	373.1\\ 	   
8.9646&	0.3144&	--143.4&	373.7\\ 	   
8.9692&	0.3176&	--140.5&	373.4\\ 	   
8.9739&	0.3208&	--139.3&	360.9\\ 	   
8.9781&	0.3237&	--146.2&	371.8\\ 	   
8.9819&	0.3263&	--138.8&	371.8\\ 	   
8.9894&	0.3315&	--136.6&	361.2\\ 	   
8.9958&	0.3359&	--140.7&	362.5\\ 	   
9.0086&	0.3447&	--151.3&	362.0\\ 	   
9.0173&	0.3507&	--141.2&	367.9\\ 	   
9.0281&	0.3581&	--139.6&	367.3\\ 	   
9.0461&	0.3705&	--135.3&	359.4\\ 	   
9.0560&	0.3773&	--134.4&	356.1\\ 	   
9.0673&	0.3850&	--127.1&	359.9\\ 	   
9.0778&	0.3923&	--123.4&	343.8\\ 	   
9.0899&	0.4006&	--124.9&	338.5\\ 	   
9.0978&	0.4060&	--118.3&	334.7\\ 	   
9.1060&	0.4117&	--112.7&	330.4\\ 	   
9.1151&	0.4179&	--113.6&	318.5\\ 	   
9.1230& 0.4235&	--110.2&	311.2\\ 	   
9.9090&	0.9638& 175.2&	--198.6\\ 	   
9.9135&	0.9668&	164.7&	--198.3\\ 	   
9.9205&	0.9716&	167.6&	--177.1\\ 	   
10.0000&	0.0263&	167.9&	--80.1\\ 	   
10.0043&	0.0293&	166.9&	--38.9\\ 	   
10.0096&	0.0329&	178.0&	--48.3\\ 	   
10.0151&	0.0367&	186.4&	--31.3\\ 	   
10.0205&	0.0404&	178.3&	--14.2\\ 	   
10.0257&	0.0440&	158.0&	--35.8\\ 	   
10.0311&	0.0477&	126.7&	17.7\\ 	   
10.0624&	0.0692&	68.4&	8.2\\ 	   
10.0727&	0.0763&	14.5&	7.8\\ 	   
10.0778&	0.0798&	37.0&	23.5\\ 	   
10.0829&	0.0833&	32.4&	30.1\\ 	   
10.9672&	0.6913&	161.3&	--226.4\\ 	   
10.9779&	0.6986&	167.6&	--219.9\\ 	   
83.803	&	0.7684	&193.9&	--298.4\\ 	   
83.831	&	0.7873	&201.9&	--314.6\\ 	   
83.865	&	0.8107	&211.0&	--333.0\\ 	   
83.880	&	0.8210	&214.4&	--339.7\\ 	   
83.896	&	0.8318	&196.5&	--303.6\\ 	   
83.911	&	0.8424	&196.0&	--302.7\\ 	   
83.926	&	0.8524	&202.7&	--316.2\\ 	   
83.941	&	0.8633	&187.4&	--285.2\\	 
\hline \\
\end{tabular}
\end{footnotesize}
\end{center}
\end{table}

\begin{figure}
\label{fig-8}
\centering
\includegraphics[height=7cm,angle=0]{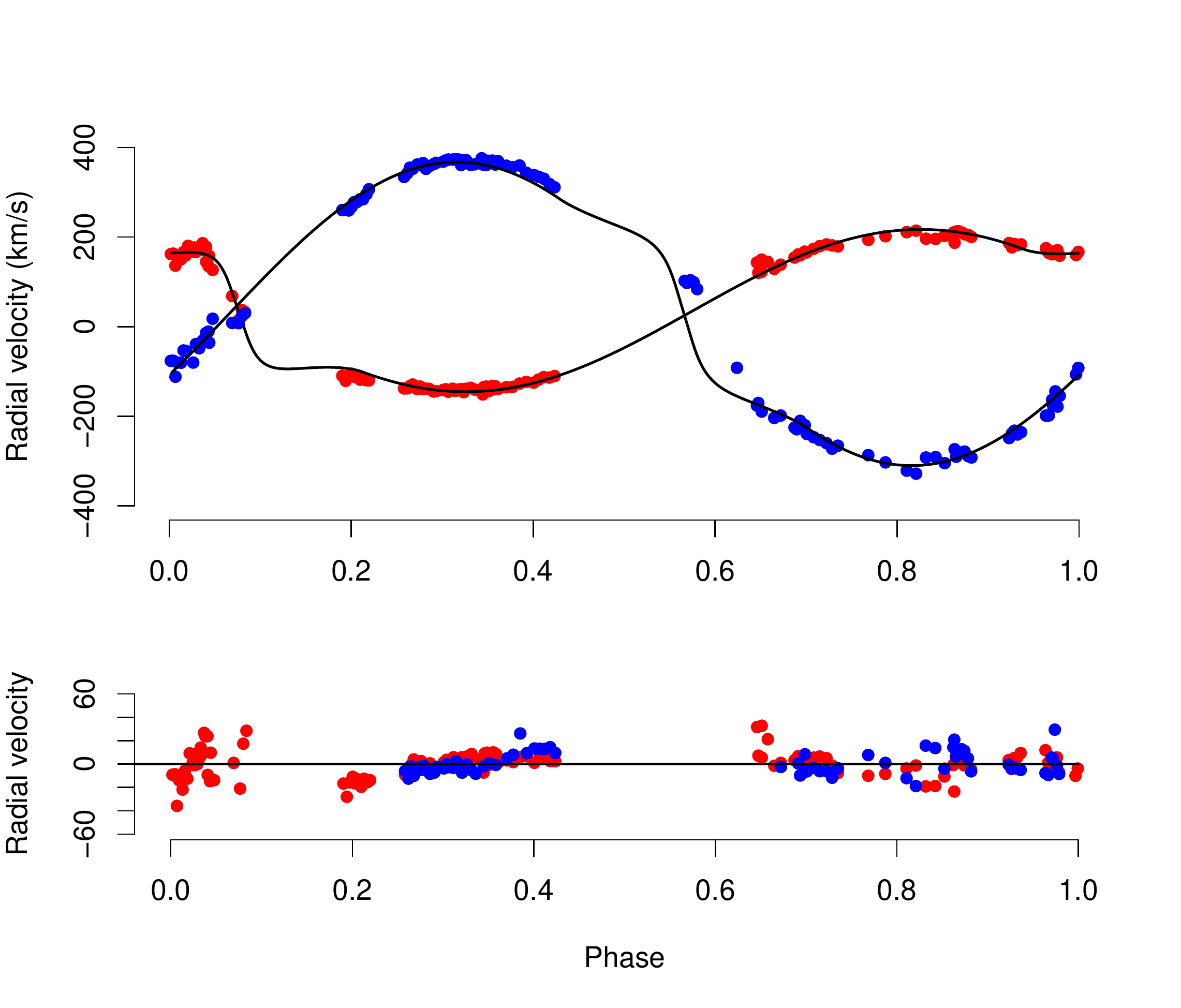}\\
\caption{Orbital RV variation of the close binary V Pup (upper panel)
with residuals (lower panel),
(red for the primary, blue for the secondary).
Full circles correspond to the UCMJO data; open circles to the 
data from the Saesteorra Observatory.
}
\end{figure}
 
\begin{table}
\begin{center}
\caption{Out-of-eclipse curve fitting results for
measured MJUO RVs for the close binary system V Pup. 
Symbols have their usual meanings.
\label{tbl-6a}} 
\begin{tabular}{lll}
\hline 
\multicolumn{1}{c}{Parameter}  & \multicolumn{1}{c}{Value}  & 
\multicolumn{1}{c}{Error}\\
\hline \\
$K_1 \sin i$ (km s$^{-1}$) & 175.4  & 3.2\\
$K_2 \sin i$ (km s$^{-1}$) & 338.8  & 5.4\\
$\Delta \phi_0$ (deg)             & 23.6   & 1.2  \\
$V_\gamma$ (km s$^{-1}$)     & 28.9 &  1.5\\
$\nu$     &  74 (p) \,\, 63 (s)  &   \\
$\chi^2/\nu$ & 1.18 (p) \,\, 1.05  & \\
$\sigma $ (km s$^{-1}$) & 7(p) \,\, 9(s)  & \\
\hline
\end{tabular}
\end{center}
\end{table}

The binary's semi-major axis $a$ can be determined from the
RV amplitudes using the inclination 
obtained from the photometric analysis.
We have,
\begin{equation}
a = \frac{P(K_1 + K_2) \sqrt{1 - e^2}}{ 2\pi \sin i}  \,\,\,  ,
\end{equation}   
where symbols have their usual meanings (Smart, 1960).
 On substitution of the foregoing values for the Eqn~1 period $P$ (in seconds) and
 the Table~6 RV amplitudes $K_1$ and $K_2$, and adopting the eccentricity 
 $e$ to be negligible, we find $a \approx 1.043 \times 10^7$ km,
 or 14.96 R$_{\odot}$.  Using the relative radii from Table~2 
 the mean radii turn out to be 5.48 and 4.59 R$_{\odot}$.
 
\subsection{Rotational velocities}

 The profiles of the He I lines, particularly the 6678 feature,
are well-defined and capable of yielding rotational and turbulence parameters.
 We show the results of such profile fitting in Table~7 and Fig~9.
\begin{figure}
\label{fig-9}
\includegraphics[height=6cm,angle=0]{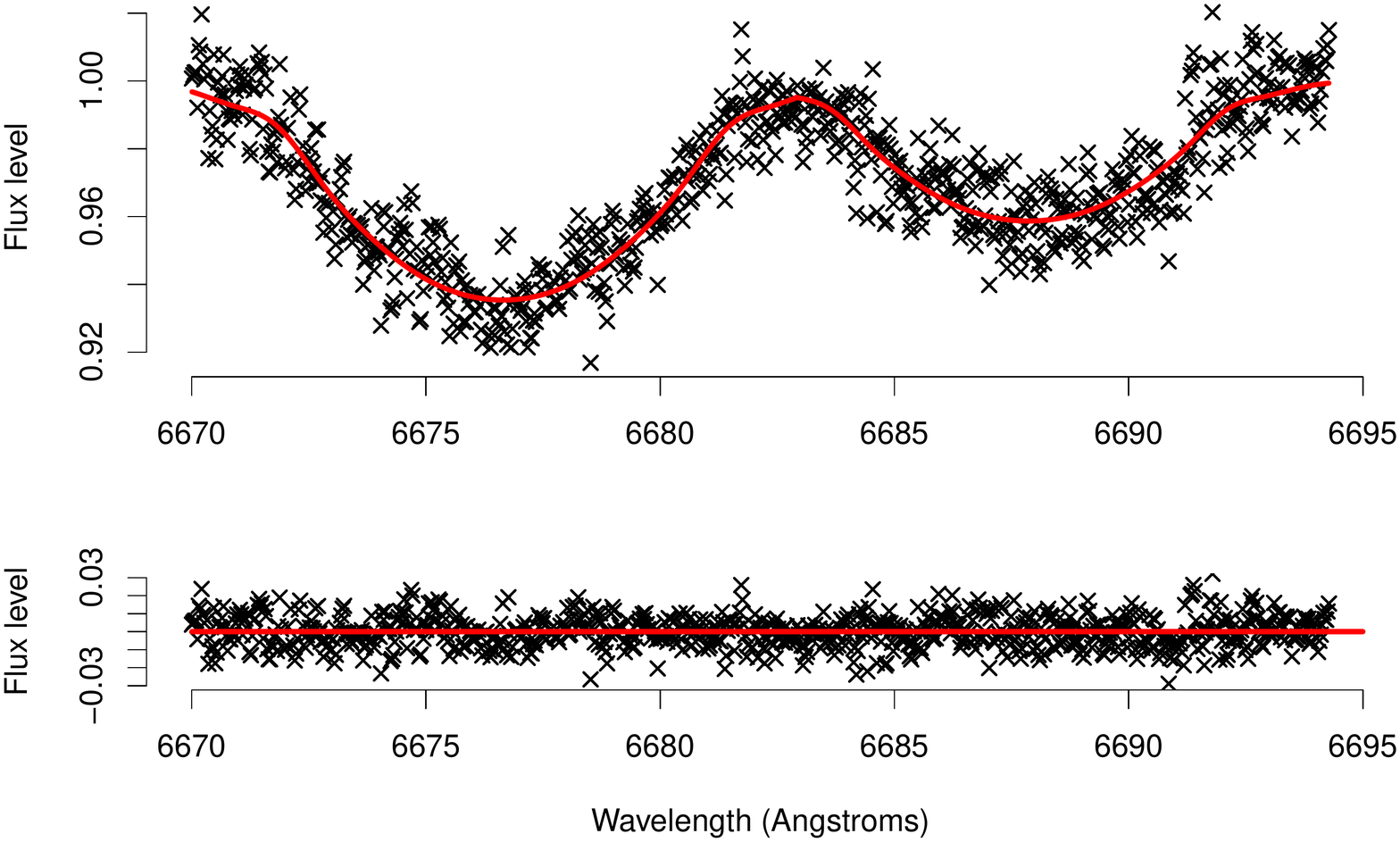}\\
\includegraphics[height=6cm,angle=0]{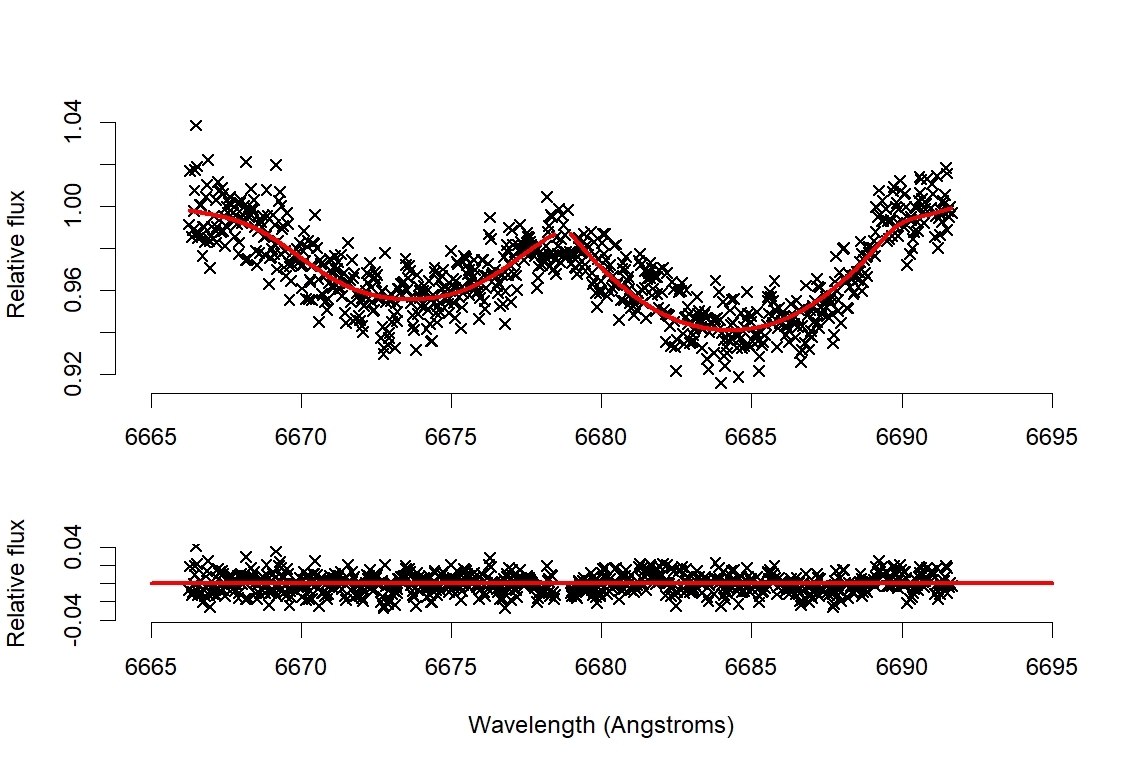}\\
\caption{Profile fitting to the HeI $\lambda$6678 lines: (upper pair)
1st elongation (primary approaching), with residuals shown beneath; 
(lower pair) 2nd elongation(primary receding), residuals below.
The lines have been fitted separately and do not quite return to the continuum level
between the absorption features: the fitting thus shows a discontinuity in that region.
}
\end{figure}

 \begin{table}
\begin{center}
\caption{Line-modelling parameters for V Pup at elongation.  
\label{tbl-6}} 
\begin{tabular}{lccc}
   \hline  \\
\multicolumn{1}{c}{Parameter}  & \multicolumn{1}{c}{1st elong.} & \multicolumn{1}{c}{2nd elong.}
& \multicolumn{1}{c}{Error}\\
\hline \\
$I_{0, 1}$   & --0.059 &--0.047 &0.002\\
$I_{0, 2}$   & --0.040 &--0.036 &0.002\\
$\lambda_1$ & 6676.78 & 6684.09&0.06 \\
$\lambda_2$ & 6687.89 &6673.69 &0.09 \\
$v_{\rm rot, 1}$ km s $^{-1}$& 218 &246 &5 \\
$v_{\rm rot, 2}$ km s $^{-1}$& 187 &209 &6 \\
$s_1$ km s$^{-1}$& 8  & 7 &2 \\
$s_2$ km s$^{-1}$& 17  & 12 &5 \\
$\Delta f$   & 0.015 &0.015 &\\
$\chi^2/\nu$ & 1.2 & 1.3&\\
\hline
\end{tabular}
\end{center}
\end{table} 
 
 The entries in Table~7 follow a similar arrangement to the other tables
 of model-fitting in this paper.
 The quantities $I_0$ correspond to the central depths on the same relative scale. 
 Mean wavelengths $\lambda_{1,2}$ are self-explanatory.
 The rotational velocities $v_{\rm rot, 1,2}$  are calculated with an
 inclination $i = 80.5^{\circ}$ (Table~2) to apply a correction for the projection.
The parameters $s_{1,2}$ represent the widths of the gaussian broadening
that is convolved with the rotational broadening.  These may be interpreted
as a measure of the turbulence in the source.  The secondary appears 
relatively more disturbed in this way than the primary, perhaps
reflecting its Roche lobe instability condition.
 
 The angular velocity of the system, using the period given in Section~2, is
 5.00$\times 10^{-5}$ radians s$^{-1}$.  With the stellar radii 
 given in the previous section, mean synchronized rotation speeds ($\omega R$)
 then turn out to be 191 and 160 km s$^{-1}$. 
These speeds are appreciably less than the measured values in Table~7,
though they are in the same ratio ($\sim$0.84), implying an
 excess of the equatorial speed above synchronism of about 15\%.  This
is comparable to the distortions from sphericity  of these rapidly rotating spheroids
(Fig~3).  Theoretical timescales for synchronization ($\sim$1 Myr; Zahn, 1977) 
 are short enough to expect a synchronized configuration.  

Rossiter effects can be seen in the data, but the measures for the secondary are
too poor (mainly as a result of blending) to be included in effective fitting trials.
It was intended to check the rotational velocity of the primary by
fitting the radial velocity curve in the eclipse region, since the
distortion from the more nearly sinusoidal variation through the 
uneclipsed phases shows a clearly visible effect in Fig~8.
This proved ineffective, however, as the difference between the results of setting
$v_{{\rm rot},1}$(sync) and $v_{{\rm rot},1}$(2$\times$sync) on the Rossiter effect
is barely noticeable against the scatter in the relevant phase range (0.0 to 0.1 in Fig~8).

\section{Absolute parameters} 
 \begin{table}
\begin{center}
\caption{Absolute parameters of V Pup system
from combined results of data fittings.  
The column of comparison numbers come from the work of Andersen et al.\ (1983).
There is closer agreement between the parameters of the present paper and those of 
the comparison than with the other papers cited by Andersen et al., though the formal errors
  given by the latter for the radii and masses are more stringent than ours (see text).
  On the other hand, our estimates of the  $V$ magnitudes,   
  given with lower uncertainties, produce a significantly closer photometric parallax to the
  independent HIPPARCOS value.
\label{tbl-7}} 
\begin{tabular}{lcccc}
   \hline  \\
\multicolumn{1}{c}{Parameter}  & \multicolumn{1}{c}{Value} & 
\multicolumn{1}{c}{Error} &\multicolumn{1}{c}{Comparison}&\multicolumn{1}{c}{Error} \\
\hline \\
$P$ d & 1.4544859 & ---   &   ---&  ---      \\
$a$ (R$_{\odot}$) & 14.96 & 0.2 & 15.27& 0.09  \\
$R_1 {\odot}$ & 5.48 & 0.18& 6.18&   0.07  \\
$R_2 {\odot}$ & 4.59 & 0.15& 4.90&   0.05 \\
$T_1$ K & 26000 & 1000& 28200&1000  \\
$T_2$ K & 24000 & 1000& 26600&1000  \\
$M_1 {\odot}$ & 14.0  &  0.5& 14.86& 0.24   \\
$M_2 {\odot}$ & 7.3 & 0.3 & 7.76& 0.14   \\
$\log L_1 {\odot}$ & 4.10 & 0.08 & 4.34& 0.06 \\
$\log L_2 {\odot}$ & 3.81 & 0.08 & 4.04& 0.07
 \\
$V_1$ mag   &  5.10 & 0.05& 4.83& 0.16  \\
$V_2$ mag   &  5.59 & 0.07& 5.50& 0.16   \\
$V_{\gamma}$ (km/s) & 20.0 & 1.1& 13.2& 2.5\\
distance pc & 320 & 10& 380& ---   30 ---     \\
age Myr &  5 & 0.8 & &\\
\hline
\end{tabular}
\end{center}
\end{table}
  
The foregoing orbital semi-major axis and radii of components are listed in Table~8.
Using the HIPPARCOS period (0.0039822 yr) with the value of $a$ as 0.06961 AU,
 we find the sum of the close binary components' masses  
directly from Kepler's third law to be 21.27 M$_{\odot}$.
Given the mass ratio $q = K_1/K_2$ = 0.518, we then find the listed individual masses,
which are sensitive to the derived
RV amplitude and inclination parameters.     
The absolute radii, $R_{1,2}$ that
come from multiplying the relative radii of Table~2 by $a$,
are determined with higher precision than the masses.
These radii, together with the temperatures from Table~2, 
produce  the luminosities given in Table~8.

\begin{figure}
\label{fig-10}
\centering
\includegraphics[height=6cm,angle=0]{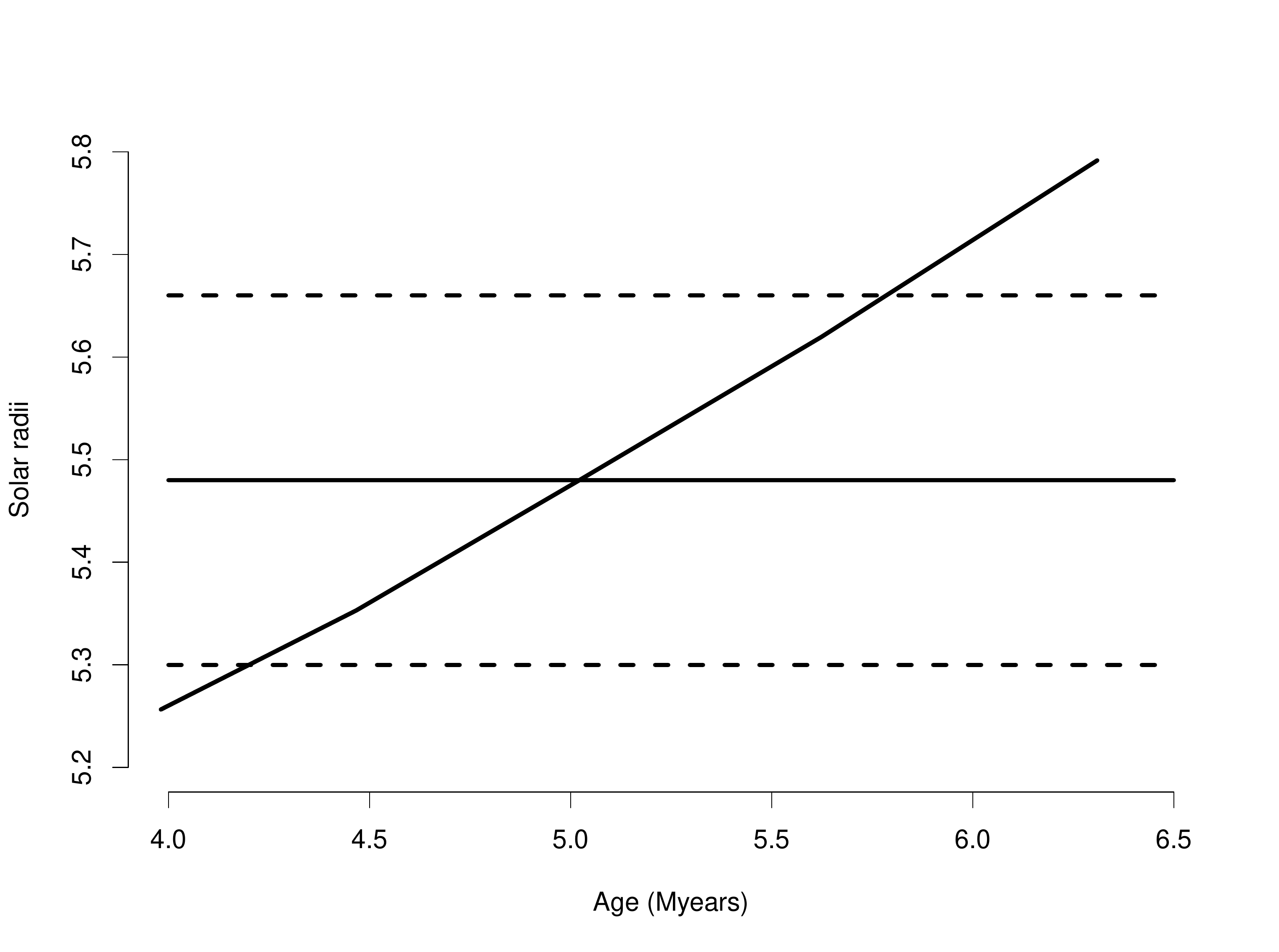}\\
\caption{Increasing radius of a 14 M$_{\odot}$ star is shown
as a function of age from the Padova models (Marigo et al., 2017),
together with the determined radius (horizontal line)
and its uncertainty limits (dotted).
An age of about 5 Myr is deduced. 
}
\end{figure} 

The set of comparison parameters  in Table~8  come from the work of Andersen et al.\ (1983).
 There is a good general agreement between our derived parameters and 
 and those of Andersen et al., though the formal errors
  given by the latter for the radii and masses are  about twice as low as 
  ours.  The differences on the mass values come mainly from the
  analysis of the radial velocity curves.  In Section~3 we gave reasons why 
  individual RV measures should have errors on the order of 10 km s$^{-1}$,
  and this would also appear to apply to the measures listed by Andersen et al.\
  (their Table~1),
  where the average datum error is close to 10 km s$^{-1}$.
  Andersen et al.\ (1983) corrected their initial RV measures according to the prescription
  of a selected model.  It is seen in  Table~3 of Andersen et al.\ (1983)
  that the differences between their initial and corrected model results 
  are significantly greater than the formal errors of the modelling parameters, while
  our radial velocity amplitudes (Table~6)
  are between the two parameter sets produced by Andersen et al.
  An empirical stance with regard to deviations of data from modelling predictions then
  allows that the real errors of parameters may be appreciably greater than the formal ones
  of a particular modelling result. Our more conservative assessment of the
  modelling accuracy comes after various curve-fitting experiments with
  different assumptions on the non-adjustable parameters.

The photometric parallax $\pi$ is derived from the formula (Budding \& Demircan, 2007;
Eqn 3.42)
\begin{equation}
\log \pi = 7.454 - \log R - 0.2V - 2{F^{\prime}}_V  \,\,\,  ,
\end{equation}
where $R$ is the absolute radius, $V$ the visual magnitude and
 ${F^{\prime}}_V$ the flux parameter (= $\log T_e - 0.1$BC;
 $T_e$ being the effective temperature and BC the corresponding bolometric correction).
The reciprocal of $\pi$ would give the distance $\rho$ to be 360 $\pm$ 12 pc,
but Eqn~3 refers to a $V$ magnitude without interstellar extinction.
The adopted $B - V$ colour of --0.17 for the components (Ducati, 2002), when compared with 
the representative unreddened colour of --0.27 for this pair of early B-type stars  
(Budding \& Demircan, 2007, Table~3.6) yields a colour excess of 0.10.  
From this, we can estimate an 
interstellar absorption value of $A_V = 0.30$ (Cardelli et al., 1989).
The effect of the absorption implies the distance would be reduced 
 by $\sim$11\%, leading to the value listed in Table~8, which is in keeping with the
derived scale of interstellar extinction for a near-Disk source.
This distance is within the error limit of the 
value given by HIPPARCOS (295 $\pm$ 30 pc).

Table~8 also gives and age estimate, that is determined from matching the calculated
radius of the MS component with Padova evolutionary models (Fig~10).  This
result is discussed further in Section~6.

\section{O -- C effects} 
 
 \begin{figure}
\label{fig-11}
\includegraphics[height=6cm,angle=0]{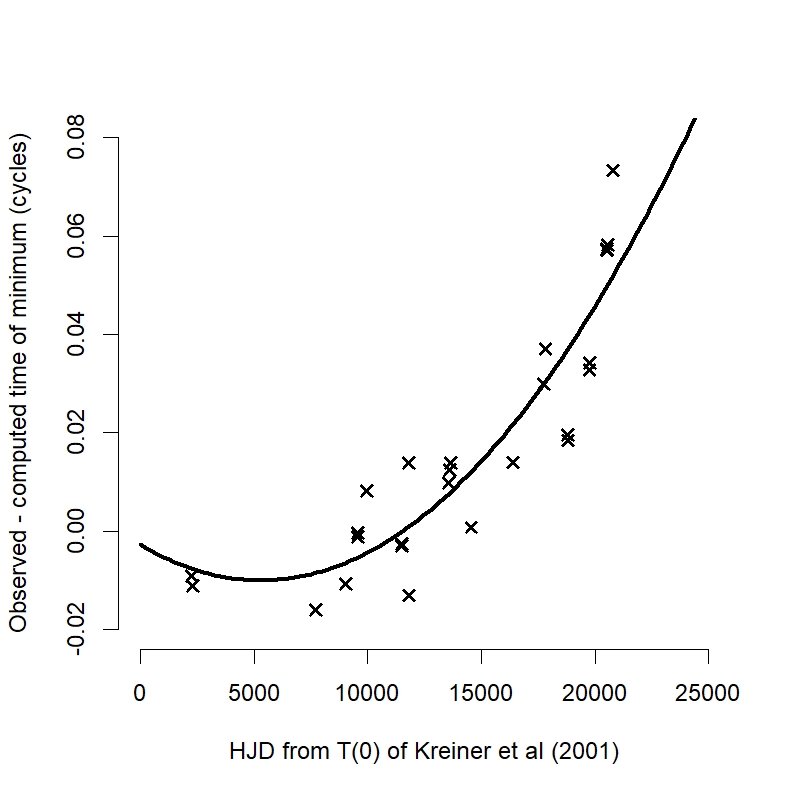}\\
\caption{Parabolic trend in the O -- C eclipse timings
of V Pup using the ephemeris of Kreiner et al.\ (2000).  
}
\end{figure} 
 
 If we plot the difference between observed times of minima for V Pup
 and those predicted by the reference ephemeris of Kreiner et al.\ (2000) (Fig~11), i.e.\
\begin{equation}
{\rm Min: I} = {\rm HJD} 2428648.2837 + 1.45448686E \,\,\,  ,
\end{equation} 
 we can see directly the general parabolic trend pointed out by Qian et al.\ (2008).  
 It is worth noting the O-C Gateway revision 
 however\footnote{ http://var2.astro.cz/ocgate/ocgate.php?star=V+Pup
 
  {\&}submit=Submit{\&}lang=en}, 
 where a longer mean period also gives a reasonable fit if we admit somewhat 
 larger timing uncertainties. 
 Our procedure is a little different to that of Qian et al.,
 in that we have added some more recent times of minima and
 separated out the parabolic trend before addressing the residuals, but our 
 result is fairly similar.  We find a second order term in the
 epoch number $E$ of order 3.5$\times 10^{-10}$, a slight reduction
 in the mean period from that given by Kreiner et al.\ (2000) of around
  5$\times 10^{-6}$ and a slight positive shift in the reference epoch of about 10 minutes.
  The uncertainties in these parameters are on the order of 10\% of their values.
  The implied rate of period variation is $\dot {P} \approx  1.8\times 10^{-7}$ d yr$^{-1}$;
  slightly less than that found by Qian et al., but within the error measures
  of the determinations.
  
  From a close examination of the residuals from the parabolic fitting,
  Qian et al. (2008) found an additional sinusoidal  variation as a 2.5$\sigma$  effect,
  with a period of $\sim$1997 d. We can see this effect clearly
  in Fig~12 for the first 17 points discussed by Qian et al.
However, after adding in the more recent data from Kreiner{\footnote{ 
https://www.aavso.org/bob-nelsons-o-c-files}; Skarka (Ho\u{n}kov\'{a} et al.\ 2013); 
the BRITE collaboration (Popowicz et al., 2017) and TESS (Ricker et al., 2015); the fit significantly deteriorates.  
These times of minima have been listed  by MB on the VSS 
website\footnote{https://www.variablestarssouth.org/budding-et-al-v-puppis/}.
According to the ephemeris of Kreiner et al.\ (2000), given as 
Eqn~4, the mean time of inferior conjunction for the radial velocity curve
shown in Fig~8 should have occurred at HJD 2458888.520 --- the 20791st such
conjunction after the reference epoch.  As can be seen in Fig~8, the conjunction 
was a little later than this, by 0.107 $\pm$ 0.007 d (0.073 cycles).  This has been 
added to Fig~11 as the rightmost point. 
  
 \begin{figure}
\label{fig-12}
\includegraphics[height=6cm,angle=0]{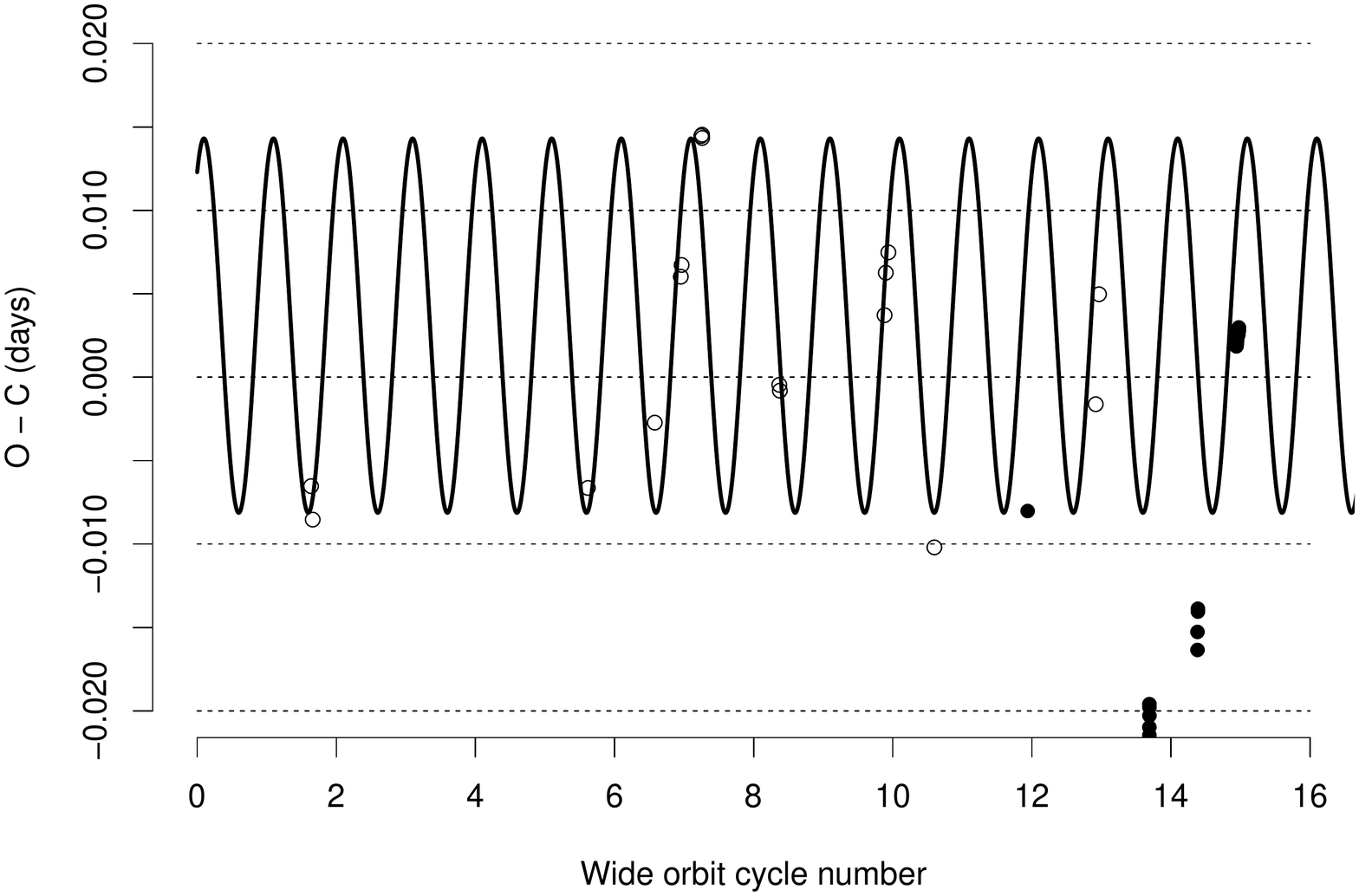}\\
\caption{Sinusoidal effect in the O -- C eclipse residuals
after removing the parabolic trend in Fig~11.  
17 selected points (open circles) conform well
to  the wide orbit period (5.47 yr) of Qian et al.\ (2008).
Later values (filled circles), however, include some 
distinct discrepancies with the sinusoidal model (see text). 
}
\end{figure}

Let us assume that the times of minimum number $E$, that is
 $T(E)$, follow the parabolic form
\begin{equation}
T(E) = A + BE + C E^2
\end{equation}
The period at $E$, usually given in days, is then given by the uniformly increasing slope
\begin{equation}
P(E) = B + 2C E  \,\,\,  ,
\end{equation}
where, in practice, $C$ is very small compared to $B$.
 
After selecting some reference epoch  $T_0$
where   $E = E_0 = 0$, the O -- C diagram  is formed
from the difference between $T(E)$ and 
corresponding $T$-values calculated along the line $T_c = P_c E$ $+$ {const.},
where $P_c = B_c$, say, and 
the constant term in the expression for $T_c$ is $A_c$, say.
The constants $A_c$, $B_c$ would be assigned to give the best position
of the line to match the observed trend in the given $E,T$ interval.  
We would expect them to be close to $A$ and $B$ in practice.  The O -- C values are then
\begin{equation}
{\rm O - C} = T(E) - T_c = (A - A_c) + (B - B_c)E + CE^2  \,\,\,  ,
\end{equation}
in days.
This is again of parabolic form, and with the same coefficient of the $E^2$ term.
Reasonably measurable values of $C$, for an O -- C of order 0.1 d or greater,
involve a suitably covered interval with a half-range for $E$ reaching to $\sim$ 10000 periods,
i.e.\ $C$ should be typically greater than about 10$^{-9}$ for a very confident 
determination.

The change of period $\Delta P$ with each event ($\Delta E = 1$) becomes $dP/dE = 2C$. 
and the rate of change of period per year at $E$ is then $730.5\, C/P(E)$.
 The amount of mass lost per year
$\Delta M_1$, follows as (Kreiner \& Zi\'{o}{\l}kowski, 1978),
\begin{equation}
\Delta{M_1} = 243.3 \frac{ M_1 C}{g(x)P^2} \,\,\,  {\rm yr}^{-1} \,\, ,
\end{equation}
where $g(x)$ is a function of the mass ratio $x  = M_1/(M_1 + M_2)$ that 
depends on what happens to the matter, as a whole,  shed by the mass-losing star,
here identified as $M_1$ -- the originally more massive component.
In the case where total mass and angular momentum of the binary are conserved 
$g(x)$ can be shown to equal $(2x -1)/(1 - x)$; so its value would be currently 
close to  --0.5, from the data in Table~8.   
 We then derive a mass transfer rate of about 
 $6\times 10^{-7}$ solar masses per year.  
 If most of the mass is lost from the system, then at low $M_1$, $g(x) \rightarrow -1$;
 so a lower mass loss rate, say $\sim 3\times 10^{-7}$ solar masses per year,
 would achieve the same period increase (i.e.\ a lower $C$ in Eqn~8).
 
 In fact, we can deduce that, if the age of the system is only $\sim$5 Myr and the erstwhile
 primary must have been bigger than half the current mass of the system,
 the average rate of mass loss of that star must be at least $\sim$10$^{-6}$ M$_{\odot}$ 
 yr$^{-1}$, and probably more than that, in keeping with the
 lack of appreciable separation of the stars that would have occurred
 with the decline to a mass ratio of 0.5 in a conservative regime. 
    
We can compare the mass loss rate with what follows from 
the mass loss formula
(cf.\ Awadalla \& Budding, 1982; Murad \& Budding, 1984):
\begin{equation}
\dot{M_1} \approx -3\eta s M_1/R_1 
\end{equation}
where we can take the mass-losing star's mass and radius ($M_1$ and $R_1$)
from (Table~8) and give consideration to the relative density parameter $\eta$ and 
rate of surface expansion $s$.
  The value $s$ might be considered 
on the basis of a Case A mass-transfer model, 
in which the loser remains in hydrostatic equilibrium
as a Main Sequence star but with a modified evolution timescale.
That would allow it to be compared with a 7.3 solar mass star of age close to 28 Myr.
The corresponding value of $s$ then 
works out as $1.0\times 10^{-7}$ solar radii per year,
using the Padova evolution models  (Marigo et al., 2017). Eqn(9) then yields 
\begin{equation}
\dot{M_1} \approx -5\eta \times 10^{-7} 
\end{equation}
solar masses per yr.
The parameter $\eta =  \rho_s/\bar{\rho}$:
 the ratio of the density in the subphotospheric layer
 where the predominating systematic motion is the outward expansion,
 to the mean density of the star as a whole,
appears to be relatively small.  Using continuity of the flow, we can write 
$\rho_s = \rho_{R1} v_{R1} \epsilon^2/s$, where $v_{R1}$ is the velocity of sound in
the atmospheric layers where the mean density is $\rho_{R1}$, and 
$\epsilon = \Omega a/ v_{R1} = h/2R_1 \approx 1/100$, $\Omega$ being the 
system's angular velocity and $h$ the mass-transferring
stream's width (cf.\ Lubow \& Shu, 1975). 
With  $v_{R1}$ estimated at $6 \times 10^{-9}$ (Cox, 2000), 
we find $\eta \approx 0.03$.  
 
 In other words,  if this Case A mass loser were to
 behave like a Main Sequence star of the same mass and luminosity,
 the calculated mass loss rate  
 is more than an order of magnitude too low to account for the 
 apparent rate of period increase.
The star is required to expand much more rapidly than that.

Of course, this follows  if we take into account an originally 
high mass of the mass-losing star.  That would
entail the earlier development of a more massive core than would
correspond with a star that remained at 7 M$_{\odot}$.
Also, $\eta$ is larger at earlier stages of interactive evolution,
as can be seen from the $\epsilon$ parameter, in the Lubow and Shu (1975) discussion,
being greater as a result of both increased mass and proximity.
If we adopt $\eta = 0.1$ and use the Padova evolution tables with 
an original value of $M_1$ of around 15 M$_{\odot}$, we find 
an expansion approaching $ 1.0\times 10^{-6}$ solar radii per year 
within the time-frame of a few million years.
 Eqn (9) then allows a
fair agreement between the amount of mass lost by $M_1$ and a Case A model
for the binary evolution.

\section{Conclusions}
 
  The V Pup system has been long recognized as a benchmark example 
  for studies of massive close binary Algol-type evolution.  It may be compared with 
  $\mu^1$ Sco, where Budding et al.\ (2015) found that the primary was 
  fairly consistent with the properties of a single star of its age and mass.  In turn, 
  that implies that the matter shed by the over-luminous secondary was probably mostly
  ejected from the system rather than transferred, thus taking angular momentum
  out of the orbit and keeping the pair in relative close proximity.  This also appears 
  to be the case for V Pup.
  This is then the regular Case A situation 
  discussed by Andersen et al.\ (1983), whose Fig~6
  shows the primary close to the trend of the Main Sequence for young massive stars
  as quantified by Popper (1980), while the secondary is significantly
  more luminous than an MS star would be at its present mass  of $\sim$7 M$_{\odot}$.  
  Although  our model is $\sim$10\% less massive than that of Andersen et al.\ the   
  parameters are essentially similar.
  
  We support this view from  the results of the present paper in two ways.
  Firstly, the photometric fittings of Section~1 show the system to be 
  consistent with a secondary star that is near to contact with its surrounding Roche lobe,
  i.e.\ an Algol configuration.  It is difficult to establish a clear 
  picture for the original arrangement of this massive system, but an interaction
  that preserves the angular momentum of the orbit would require the
  star centres to be implausibly near to each other at around 1 solar radius
  when at their closest.  The generally adopted scenario thus implies that matter shed
  by the loser is mostly driven out of the system by intense radiation pressure,
  taking with it a fair proportion of the original systemic angular momentum.
  If about half the system's angular momentum was lost in this way,
  an original pair of near equal masses totalling around 30 M$_{\odot}$
  would have had a separation of order 10 solar radii,
  the two near-contact components having mean radii of about 0.4 of this.   

  The Padova modelling  (Marigo et al., 2017) giving rise to Fig~10 shows that 
  a ZAMS star of 15 M$_{\odot}$ would be expanding at a rate of $\sim$0.5 solar radii
  per Myr; in other words, not many Myr would elapse before the
  more massive star encounters its surface of limiting stability and
  starts to lose matter to its surroundings.  
     
  Independent confirmation  of this scenario comes from Fig~6, where the position of
  V Pup's primary star in the luminosity-temperature diagram places it 
  at about 1/3 of the way across its Main Sequence track and just entering
  the instability strip corresponding to massive early type stars that are
  associated with the $\beta$ Cep pulsational effect.
  This primary star, then, appears to be behaving much  
  as a single Population I star would behave
  at its deduced age and mass. 
  We  believe that this is the fourth $\beta$ cepheid variable to be reported
  having a precise mass measurement
  after V453 Cyg A (Southworth et al., 2020), 
  one component of CW Cep  (Lee \& Hong, 2010), 
   and VV Ori (Southworth, 2021). 
  A preliminary exploration of the pulsation spectrum was carried out,
  giving support to the likely nature of the low-amplitude variability,
  but it is likely that the power spectrum is affected by components
  associated with submultiples of the close binary period.  This subject
  is rather outside the scope of the present article, but
  further study of the bright object V Pup is clearly called for in this connection.}
  
  The ongoing mass-loss of the secondary does raise the issue of detectable period
  variation, and we have confirmed the general trend of period increase reported 
  by Qian et al.\ (2008), though with about a 15\% lower rate.
  We could not confirm the 5.47 yr additional cyclical variation found by Qian et al.\
  after admitting more new times of minima (ToMs),
  though there remain relatively large discrepancies between individual ToMs 
  and the longer term parabolic trend, particularly in the  ToMs
  reported by Skarka (Ho\u{n}kov\'{a} et al.\ 2013).   It is possible that some periodic term in the 
  calculated times of minima
  may  reduce these discrepancies, but we could not establish 
  this with a high  degree of confidence.
  
In this connection we note that we found an improvement in the quality of 
fit to the TESS data could be achieved by the inclusion of a small
amount ($\sim$10\%) of third light. The point can be readily ascertained
by a direct comparison of the eclipse amplitudes in Figs~1 and 2;
the longer-wavelength window of TESS allowing a greater proportion
of light from a somewhat cooler companion to diminish the relative proportions
of both eclipses. Eaton (1978) also found a third light contribution,
of order $\sim$3\% in the V range. 
 A mid-to-late B type MS third star would meet the implied relative
 luminosity.  We could not find any clear evidence of this star
 in the spectral data, however; though this is unsurprising
 given the nature of the early type rapidly rotating stars involved and
 their relative light levels in the optical range.  The question of a third major body in
the V Puppis system thus appears still unresolved and 
awaits further more detailed investigation.

\section{Acknowledgments}
 
It is  a pleasure to express our appreciation of the high-quality and 
ready availability, via the Mikulski Archive for Space Telescopes (MAST),
of data collected by the TESS mission.
 Funding for the TESS mission is provided by the NASA Science Mission Directorate
 
Generous allocations of time on the 1m McLellan Telescope and HERCULES spectrograph at the University of Canterbury Mt John Observatory in support of the
Southern Binaries Programme have been made available through its TAC and supported by its  Director, Dr.\ K.\ Pollard and previous Director, Prof.\ J.B.\ Hearnshaw.  Useful help at the telescope was provided by the UCMJO management (N.\ Frost)
assisted by F.\ Gunn.   Considerable assistance with our use of the {\sc hrsp} software 
package was given by its author Dr.\ J.\ Skuljan. 

Encouragement and support for this programme has been shown by the 
the School of Chemical and Physical Sciences
of the Victoria University of Wellington, as well as the Royal Astronomical
Society of New Zealand and its Variable Stars South section
(http://www.variablestarssouth.org). 
Further details on the photometry supporting the Southern Binaries Programme
of the VSS are available from its Director Mark Blackford. 

\section{Data availability}
All data included in this article are available as listed in the paper or
from the online supplementary material it refers to. 

{}


\begin{thebibliography}{}
 
\bibitem[\protect\citeauthoryear{Andersen}{1983}]{andersen}
Andersen, J., Clausen, J.V., Gimenez, A., Nordstrom, B.,  1983, A\&A, 128, 17

\bibitem[\protect\citeauthoryear{Awadalla}{1982}]{awad}
Awadalla, N.\ S., Budding, E., 1982 
{\em Binary and multiple stars as tracers of stellar evolution: Proc.\ 69 IAU Coll.}, (Reidel), 1982, p.\ 239

\bibitem[\protect\citeauthoryear{bahcall}{1975}]{bahcall}
Bahcall, J.\ N., Charles, P.\ A., Davison, P.\ J.\ N., Stanford, P.\ W., Kellogg, E., York, D., 1975, MNRAS, 171, 41

\bibitem[\protect\citeauthoryear{Budding}{1990}]{bank}
Banks, T., Budding, E., 1990, ApSS, 167, 221

\bibitem[\protect\citeauthoryear{bell}{1987a}]{bell1}
Bell, S.\ A., Adamson, A.\ J.,  Hilditch, R.\ W., 1987a, MNRAS, 224, 649
 
\bibitem[\protect\citeauthoryear{bell}{1987b}]{bell2}
Bell, S.\ A., Kilkenny, D.,   Malcolm, G.\ J., 1987b, MNRAS, 226, 879

\bibitem[\protect\citeauthoryear{Blackford}{2019}]{bla}
Blackford, M.\ G., Erdem, A., S\"{u}rgit, D., \"{O}zkarde\c{s}, B., Budding, E., Butland, R., Demircan, O., 
2019, MNRAS, 487, 161

\bibitem[\protect\citeauthoryear{Budding}{2008}]{bud1}
Budding E., 
2008, {\em APRIM 2008: Proceedings of the 10th Asian-Pacific Regional IAU Meeting
Kunming, China, August 3-6, 2008},
Eds.\ Shuang Nan Zhang, Yan Li,  Qing Juan Yu,
National Observatories of China Press, 33

\bibitem[\protect\citeauthoryear{Budding}{2007}]{bud3}
Budding E., Demircan O., 2007, {\em An Introduction to Astronomical
Photometry}, Cambridge Univ.\ Press 

\bibitem[\protect\citeauthoryear{Budding}{2015}]{bud4}
Budding, E., Butland, R., Blackford, M.\ G., 2015, MNRAS, 448, 3784
 
\bibitem[\protect\citeauthoryear{Cardelli}{1989}]{car}
Cardelli, J.\ A., Clayton, G.\ C., Mathis, J.\ S., 1989,
ApJ, 345, 245

\bibitem[\protect\citeauthoryear{Cester}{1977}]{ces}
Cester, B.; Fedel, B.; Giuricin, G.; Mardirossian, F.; Pucillo, M.
1977, A{\&}A, 61, 275

\bibitem[\protect\citeauthoryear{Cherepashchuk}{1976}]{che}
Cherepashchuk, A.\ M., 1976, SvAL, 2, 138

\bibitem[\protect\citeauthoryear{Chlebowski}{1991}]{chl}
Chlebowski, T., Garmany, C.\ D.,
1991, ApJ, 368, 241

\bibitem[\protect\citeauthoryear{Corcoran}{1996}]{cor}
Corcoran, M.\ F.,
1996, MxAC, 5, 54

\bibitem[\protect\citeauthoryear{cousins}{1947}]{cousins}
Cousins, A.\ W.\ J., 1947,
MNAS South Africa, 6, 91

\bibitem[\protect\citeauthoryear{Cox}{2000}]{cox}
 Cox, A.\ N., 2000, {\em Allen's Astrophysical Quantities}, p393.

\bibitem[\protect\citeauthoryear{Desnoux}2005]{den}
Desnoux, V.,  Buil, C., 2005,
{\em Soc.\ Astron.\ Sci.\ Ann.\ Symp.},  24, 129

\bibitem[\protect\citeauthoryear{Ducati}{2002}]{duc}
Ducati, J.\ R., 2002,
  {\em CDS/ADC Collection of Electronic Catalogues}, 2237, 0

\bibitem[\protect\citeauthoryear{Eaton}{1978}]{eat}
Eaton, J.\ A., 1978, Acta Astron., 28, 63

\bibitem[\protect\citeauthoryear{ESA}{1997}]{esa}
ESA, 1997, The HIPPARCOS and Tycho Catalogues, ESA SP-1200
 
\bibitem[\protect\citeauthoryear{1962}{friebos}]{friebos}
Friebos, H.\ O., 1962, Ap.J., 135, 762

\bibitem[\protect\citeauthoryear{Garcia}{1989}]{gar}
Garcia, M.\ R., Bailyn, C.\ D., Grindlay, J.\ E., Molnar, L.\ A., 1989, ApJ, 341, L75

\bibitem[\protect\citeauthoryear{giacconi}{1974}]{giacconi}
Giacconi, R., Murray, S., Gursky, H., Kellogg, E., Schreier, E., Matilsky, T., Koch, D.,  Tananbaum, H., 1974, ApJS, 27, 37
 
\bibitem[\protect\citeauthoryear{gies}{1984}]{gies}
Gies, D.\ R.,  Bolton, C.\ T., 1984, ApJ, 27, 37

\bibitem[\protect\citeauthoryear{groote}{1978}]{gro}
Groote, D., Kaufmann, J.\ P., Hunger, K., 1978, A\&A, 63, L9

\bibitem[\protect\citeauthoryear{Gummersbach}2014]{gum}
Gummersbach C., Kaufer A., 2014,
{\em Synthetic Spectra of B Main-Sequence Stars from 3000 --- 10000 \AA},  

{\footnotesize https://www.lsw.uni-heidelberg.de/projects/hot-stars 

/websynspec.php}

\bibitem[\protect\citeauthoryear{Hearnshaw}{2002}]{hea}
Hearnshaw J.\ B., Barnes S.\ I., Kershaw G.\ M., Frost N., Graham G., 
Ritchie R., Nankivell G.\ R., 
 2002, Exp.\ Astron., {13}, 59 


\bibitem[\protect\citeauthoryear{Honkova}{2013}]{hon} 
 Ho\u{n}kov\'{a}, K.\ et al.(80 authors), 2013, OEJV, 160, 1 (BRNO Cont.\ No.\ 38)

\bibitem[\protect\citeauthoryear{Idaczyk}2013]{idb}
Idaczyk R., Blackford M., Butland R., Budding E.,   
South.\ Stars, 2013, 52, No.\ 3, 16

\bibitem[\protect\citeauthoryear{Jaschek}1987]{jas}
Jaschek C., Jaschek M., 1987, {\em The Classification of Stars} CUP, Cambridge

\bibitem[\protect\citeauthoryear{Khopolov}{1987}]{kho}
Khopolov, P.\ N. et al.\ (11 authors), 1987, GCVS, Nauka 

\bibitem[\protect\citeauthoryear{koch}{1981}]{koch}
Koch, R.\ H., Bradstreet, D.\ H., Perry, P.\ M., Pfeiffer, R.\ J., 1981, PASP, 93, 621

\bibitem[\protect\citeauthoryear{Kopal}{1959}]{kop}
Kopal Z., 1959, {\em Close Binary Systems}, Chapman \& Hall, London

\bibitem[\protect\citeauthoryear{kriener}{2001}]{kriener}
Kreiner, J.\ M., Kim, C.-H., Nha, I.-S., 2001,
{\em An Atlas of O -- C Diagrams of Eclipsing Binary Stars}, 
 Cracow: Wydawnictwo Nauk., AP
 
\bibitem[\protect\citeauthoryear{kriener}{1978}]{kri2}
Kreiner, J.\ M., Zi\'{o}{\l}kowski, J., 1978, Acta A, 28, 497

\bibitem[\protect\citeauthoryear{lee}{2010}]{lee}
Lee, J.\ W., Hong, K, 2010, arXiv 2010.03711

\bibitem[\protect\citeauthoryear{Leone}{1998}]{le0}
 Leone, F., Lanzafame, A.\ C., 1998,  A{\&}A 330, 306
 
\bibitem[\protect\citeauthoryear{lubow}{1975}]{lubow}
Lubow, S.\ H., Shu, F.\ H., 1975, Ap.J, 198, 383

\bibitem[\protect\citeauthoryear{Maccarone}{2019}]{maccarone}
Maccarone, T.\ J., Fender, R.\ P., Knigge, C.,  Tziomis, A.\ K.,  2009, MNRAS, 393, 1070

\bibitem[\protect\citeauthoryear{Mann}{2015}]{man}
Mann, A.\ W., von Braun, K., 2015, PASP, 127, 102

\bibitem[\protect\citeauthoryear{Marigo}{2017}]{mar}
Marigo P., Girardi L., Bressan A., Groenewegen 
M.\ A.\ T., Silva L.,  2017, A\&A, {482}, 833

\bibitem[\protect\citeauthoryear{Maury}{1920}]{mau} 
Maury, A., 1920, HA, 84, 142

\bibitem[\protect\citeauthoryear{Murad}{1984}]{mur} 
Murad, I.\ M.; Budding, E.1984,  Observatory, 104, 83

\bibitem[\protect\citeauthoryear{nowak}{2002}]{nowak}
Nowak, M.\ A., Heinz, S., Begelman, M.\ C., 2002, Ap.J., 573, 778

 \bibitem[\protect\citeauthoryear{pamyatnykh}{2007}]{pam}
Pamyatnykh, A.\ A., 2007, CoAst, 150, 207

\bibitem[\protect\citeauthoryear{parkin}{2011}]{par}
Parkin, E.\ R., Pittard, J.\ M., Corcoran, M.\ F., Hamaguchi, K., Stevens, I.\ R., 
Gosset, E., Rauw, G., De Becker, M., 2011, Bull.\
Soc.\ Roy.\ Sci.\ Li\`{e}ge, vol.\ 80, p.\ 610;
eds.\ G.\ Rauw, M.\ De Becker, Y.\ Naz\'{e}, J-M. Vreux, P.\ Williams

\bibitem[\protect\citeauthoryear{pittard}{2009}]{pit}
Pittard, J.\ M., 2009, MNRAS, 396, 1743

\bibitem[\protect\citeauthoryear{pollock}{19987}]{pol}
Pollock, A.\ M.\ T.,  1987,  ApJ., 320, 283

\bibitem[\protect\citeauthoryear{Popowicz}{2017}]{pop} 
Popowicz, A.\ et al.\ (18 authors), 2017, A{\&A}, 605, 26

\bibitem[\protect\citeauthoryear{popper}{1947}]{pop1}
Popper, D.\ M., 1947, Obs, 67, 227

\bibitem[\protect\citeauthoryear{popper}{1980}]{pop2}
Popper, D.\ M., 1980, Ann.\ Rev. A{\&}A, 18, 115

\bibitem[\protect\citeauthoryear{Priedhorsky}{1983}]{pri}
Priedhorshy, W.\ C.,  Terrell, J., 1983, ApJ, 273, 709
 
\bibitem[\protect\citeauthoryear{Prilutski}{1976}]{prl}
Prilutski, O.\ F.,  Usov, V.\ V, 1976, SvA, 20, 2 
 
\bibitem[\protect\citeauthoryear{qian}{2008}]{qian}
Qian, S.-B., Liao, W.-P., Fernandez-Lajus, E., 2008, Ap.J., 687, 466

\bibitem[\protect\citeauthoryear{rauw}{2016}]{rau}
Rauw, G., Naz\'{e}, Y.,   2016, AdSpR, 58, 761

\bibitem[\protect\citeauthoryear{Ricker}{2015}]{ric}
Ricker, G.\ R., et al (58 authors), 2015,
JATIS, 1, id. 014003

\bibitem[\protect\citeauthoryear{Ruf}{1999}]{ruf}
Ruf, T., 1999, Biol.\ Rhythm Research, 30, 178

\bibitem[\protect\citeauthoryear{schneider}{1979}]{sch}
Schneider, D.\ P., Darland, J.\ J., Leung, K-C., 1979, AJ, 84, 236

\bibitem[\protect\citeauthoryear{Skuljan}{2004}]{sk1}
Skuljan J., 2004, {\em HRSP Version 2.3}, Univ.\ Canterbury publ., p16

\bibitem[\protect\citeauthoryear{Skuljan}{2012}]{sk2}
Skuljan J., 2012,  {\em HRSP Version 5}, private communication

\bibitem[\protect\citeauthoryear{Smart}{1960}]{sma}
Smart, W.\ M., {\em Textbook on Spherical Astronomy}, 1960,
CUP, 4th ed.

\bibitem[\protect\citeauthoryear{Southworth}{2020}]{sou1}
Southworth, J., Bowman, D.\ M., Tkachenko, A.,  Pavlovski, K., 2020,
MNRAS 497, L19

\bibitem[\protect\citeauthoryear{Southworth}{2021}]{sou2}
Southworth, J., Bowman, D.\ M., Pavlovski, K., 2021, MNRAS 501L 65

\bibitem[\protect\citeauthoryear{Stevens}{1992}]{ste}
Stevens, I.\ R., Blondin, J.\ M., Pollock, A.\ M.\ T.,  1992,  ApJ., 386, 265

\bibitem[\protect\citeauthoryear{Stickland}{1998}]{sti}
Stickland, D.\ J., Lloyd, C., Pachoulakis, I., Koch, R.\ H., 1998, 
 Obs., 118, 356

\bibitem[\protect\citeauthoryear{sysbesma}{1986}]{sysbesma}
Sybesma, C.\ H.\ B., 1986, A\&A, 159, 108
 
\bibitem[\protect\citeauthoryear{Terrell}{2005}]{ter}
Terrell, D., Munari, U., Zwitter, T., Wolf, G., 2005, MNRAS, 360, 58

\bibitem[\protect\citeauthoryear{Thizy}{2007}]{thi}
Thizy, O., 2007, {\em Soc.\ Astron.\ Sci.\ Ann.\ Symp.}, 26,  31

\bibitem[\protect\citeauthoryear{Worley}{1997}]{wor}
Worley, C.\ E., Douglass, G.\ G., 1997,
 	A{\&}A Suppl.\ Ser., 125, 523)

\bibitem[\protect\citeauthoryear{Williams}{1886}]{wil} 
Williams, A.\ S., 1886, MNRAS, 47, 91
 
\bibitem[\protect\citeauthoryear{yakut}{2005}]{yakut}
Yakut, K., Eggleton, P.\ P., 2005, Ap.J., 629, 1055


\bibitem[\protect\citeauthoryear{york}{1976}]{york}
York, D.\ G., Flannery, B.,  Bahcall, J., 1976, ApJ, 210, 143

\bibitem[\protect\citeauthoryear{Zahn}{1977}]{zah}
Zahn J-P., 1977, A{\&}A, 57, 383
 
\bibitem[\protect\citeauthoryear{Zdziarski}{2007}]{Zdziarski}
Zdziarski, A.\ A., Gierlinski, M., Wen, L., Kostrzewa, Z., 2007, MNRAS, 377, 1017

\end{thebibliography}
 \end{document}